\documentstyle[12pt]{article}
\def\half{{1 \over 2}}
\def\lrarrow{\leftrightarrow}
\def\a{\alpha}
\def\b{\beta}
\def\r{\rho}
\def\s{\sigma}
\def\v{\varepsilon}
\def\d{\delta}
\def\t{\theta}
\def\ab{\alpha\beta}
\def\p{\partial}
\def\m{\mu}
\def\n{\nu}
\def\mn{\mu\nu}
\def\g{\gamma}
\def\l{\lambda}
\def\G{\Gamma}
\def\o{\omega}
\def\O{\Omega}
\def\lan{\langle}
\def\ran{\rangle}

\begin{document}
\begin{flushright}
{ IC/90/97}
\end{flushright}
\vspace{0.4cm}

\begin{center}
{\small International Atomic Energy Agency\\
\vspace{0.2cm}

and\\
\vspace{0.2cm}

United Nations Educational Scientific and Cultural Organization}\\
\vspace{0.4cm}

INTERNATIONAL CENTRE FOR THEORETICAL PHYSICS


\vspace{1.5cm}

{\bf NONABELIAN $N=2$ SUPERSTRINGS}
\vspace{1cm}

{\small A.P. Isaev \footnote{Laboratory of Theoretical Physics,
JINR, Dubna, P.O. Box 79, Head Post Office, Moscow, USSR.}\quad
and \quad E.A. Ivanov $^{1}$

International Centre for Theoretical Physics, Trieste, Italy.}

\bigskip\bigskip
ABSTRACT
\end{center}
\vspace{0.4cm}

{\small The Green--Schwarz covariant $N=2$ superstring action can be consistently deduced as the
action of the Wess--Zumino--Witten (WZW) sigma model
defined on the direct product of two $N=1$, $D=10$ Poincar\'e
 supertranslation groups.
Generalizing this result, we construct new WZW sigma models on the supergroups with a nonabelian
even part and interpret them as models of superstrings moving on the supergroup manifolds.
We show that these models are completely integrable and in
some special cases possess fermionic $\kappa$--symmetry.}
\vspace{2cm}

\begin{center}
{MIRAMARE -- TRIESTE}
\vspace{0.2cm}

{\small April 1990}

\end{center}
\vfill

\section{INTRODUCTION}
An important class of string models is represented by strings moving in a curved background chosen
as the product of flat Minkowski space and a nonabelian group manifold \cite{[1]}. A necessity to
consider the models of that kind arises, e.g., while deducing realistic four--dimensional local
field theories as a low--energy limit of strings moving in a $d$--dimensional space--time: the
remaining $(d-4)$ dimensions of the latter are to be compactified in a proper manner (for example,
on a group space). Self--consistency of the string theory of that type requires it to be
conformally invariant \cite{[2]}. For ensuring this, the string coordinates valued in the group
manifold should be described by the conformally invariant Wess--Zumino--Witten (WZW) sigma-model
coupled to the world-sheet gravity. So, the group space string models supply nice examples of a 2D
conformal field theory solvable both on classical and quantum levels \cite{[1],[3],[4]}. It is also
worth mentioning that these models provide a way of introducing gauge degrees of freedom on the
string
\cite{[5]}.

It is tempting to construct the group space analogs of superstrings in the Green-Schwarz (GS)
covariant formulation \cite{[6]}, with nonabelian supergroup manifolds as the target ones. These
models could offer an appropriate laboratory for getting a further insight into the quantum and
algebraic structure of the covariant superstring theory.

In the present paper we propose a general method for formulating such nonabelian superstring
models (of the type II) in a
consistent way, proceeding from a few primary principles.
We confine our study here to the classical case.
Full quantum consideration will be given elsewhere.

As distinct from the case of bosonic strings, it is not so easy to define a viable nonabelian
generalization of the GS superstring. One of the reasons is that even the flat target superspace
possesses a nontrivial geometry in what concerns pure Grassmann dimensions (there are corresponding
nonzero torsion components). This gives rise to the existence of non--trivial Wess--Zumino terms
which have to be necessarily included in the GS covariant action for reducibility of the latter (in
a special gauge) to an action of a free field theory \cite{[6]}. The reducibility is ensured by
local ${\kappa}$--supersymmetry which plays the same role for the GS action as local world--sheet
conformal or superconformal symmetries for the actions of bosonic and spinning strings. Any
superstring action pretending to be a generalization of the original GS one should either inherit
all the remarkable features of the latter or obey instead some self--consistency requirements of
another kind
(e.g., such could be the property of complete classical
and quantum integrability).

What is definitely desirable to maintain is the nice interpretation of GS superstrings as WZW sigma
models associated with some superspaces as the target manifolds \cite{[7]}. However, for the type
II superstrings we deal with in the present paper this sigma model interpretation is not
straightforward even in the case of a flat target superspace. The point is that the relevant WZ
term cannot be immediately related to $10D$
$N=2$ Poincar\'e  supertranslation algebra which is usually assumed to underly the
type II GS superstring (see e,g, \cite{[8]}). In particular, this term does not possess the
automorphism $SO(2)$ symmetry inherent in the above algebra.

In \cite{[9]} we have argued that for an unambiguous construction of the WZ term the type II GS
action should be defined on the product of two $N=1$ supertranslation groups rather than on $N=2$
 supertranslation group.
Half the original bosonic group parameters is removed owing to the
special choice of the target superspace as the coset
$G\otimes G/G_+$ where $G_+$ is an abelian subgroup generated
by a linear combination of the two original even translation
generators (the product $G\otimes G$ is realized on this coset manifold
just as $N=2$
Poincar\'e supertranslation group).
The WZ term properly constructed out of the Cartan one--forms given on the product $G\otimes G$ is
precisely the one entering into the GS action.
The breaking of $SO(2)$ symmetry in this term is naturally related to the fact that the
supergroup $G\otimes G$ has no such automorphisms.

As a by--product of our interpretation, we have found a zero--curvature representation for the
classical covariant GS superstring equations in an arbitrary gauge. The zero--curvature
representation superalgebra turned out to be a sum of the two mutually commuting ones, thus
reflecting the $G\otimes G$ structure of the GS action. Later on, we extended this construction to
an arbitrary supergroup with the commutative even part \cite{[10]}. The arising generalized GS type
sigma models, under a fixed relative strength of the WZ and conventional terms in the action,
display complete intergrability and, with some further restrictions  on the supergroup structure
constants, possess local $\kappa$--supersymmetry.

The next natural step (and this is previsely what we do here) is to carry out an analogous
construction for the essentially nonabelian supergroups $G_1, G_2$
with non--commutative even parts.
The superstring models obtained in this way involve in the bosonic sector the group manifold
string models (the WZW sigma models) and thus can be viewed as a
genuine superextension of the latter.

In this paper we present the group--theoretical basics of nonabelian superstring models and
construct the invariant action for them. We begin in Sec.2 with recalling the main features of our
formulation of the ordinary Type II GS superstring. Further, in Sec.3, we extend this to the
nonabelian case. As a natural guiding principle, we impose the requirement that the corresponding
models are reduced to the GS type models considered in \cite{[10]} when the underlying nonabelian
supergroups contract into those with commutative even parts (we call the latter models abelian
superstrings). A novel point is that for ensuring the correct flat superspace limit the supergroups
$G_1$ and $G_2$ one starts with have to be dual to each other in Cartan's sense \cite{[11]}
$(G_2=G^*_1)$ rather than isomorphic as in the abelian case considered in \cite{[9],[10]}.
 {}From the group--theoretical point of view, the models which we obtain correspond
 to the nonsymmetric coset
$G_1\otimes G_2/G_+$ where $G_+$ is a diagonal subgroup
in the product of two isomorphic even subgroups of $G_1$ and
$G_2 =G^*_1$.
This is a crucial difference of this class of superstring models from, e.g. the model proposed
recently by Green \cite{[12]}. The latter is constructed as a WZW sigma--model for the principal
chiral field on a supergroup $G$ which is an extension of the Poincar\'e supertranslation group by
a spinor central charge.
 {}From the group--theoretical point of view, it corresponds to the symmetric coset
$G_L\otimes G_R/G_{diag}$.

We demonstrate that on the coset $G_1\otimes G_2/G_+$ it is possible to define a correct WZW action
and the latter is reduced just to the standard GS one in the flat superspace limit, when the
parameters of the action are adjusted in a proper way. Once again, the models constructed are
completely integrable for arbitrary gauge fixing. What concens local $\kappa$--supersymmetry, it is
present in its standard form only under a certain additional condition on the structure constants,
like in the abelian case \cite{[10]}.

In further publications we are planning to investigate the Hamiltonian structure  of our nonabelian
models (along the line of Ref. \cite{[13]})
\setcounter{equation}{0}

\section{$N=2$ GREEN--SCHWARZ SUPERSTRING
AS A WZW SIGMA MODEL ON THE PRODUCT OF TWO $N=1$
SUPERTRANSLATION GROUPS}

To make more clear the motivations for considering the superstring--like models constructed in the
next Section, it is instructive to start with recalling the basic points of our interpretation of
the $N=2$ GS superstring \cite{[9]}.

\subsection{Underlying concepts}

The major statement of \cite{[9]} is that for achieving the complete analogy between 2D WZW sigma
models and $N=2$ GS superstring
(both of IIA and IIB types) the latter has to be
constructed on the product of two $N=1$
 supertranslation groups rather than on the $N=2$ supergroup
as it has been originally proposed in \cite{[7],[8]}. This allows an algorithmic construction of
the
$N=2$
 superstring WZ term out of the Cartan 1--forms defined
 on the above product.
$N=2$ supersymmetry of the action is ensured due to a special choice of the target manifold
as a proper coset of this product of supergroups.

In more detail, the basic concepts of our construction are as follows:

A. One starts with the product of two isomorphic $N=1$, $D=10$ supertranslation groups
$G_1\otimes G_2$ generated,
respectively, by the generators
$(P^1_\mu,Q^1_\alpha)$ and $(P^2_\mu, Q^2_\alpha)$\footnote{
We use the standard $D=10$ conventions
$\Gamma^\mu_{\alpha\beta}=(\gamma^\mu)^\delta_\alpha C_{\delta\beta} =
\eta^{\mu\nu}\Gamma_{\nu,\alpha\beta}$
where
$\gamma^\mu$ are 32$\times$ 32 $D=10$ Dirac $\gamma$--matrices,
$C_{\alpha\beta} = -C_{\beta\alpha}$ is the
charge conjugate matrix and
$\eta^{\mu\nu}={\rm diag} (1,-1,-1,\dots,-1)$.}
$$\{Q^1_\a,Q^1_\b\} = -\G^\m_{\ab}P^1_\m,[P^1_\m,Q^1_\a]=
[P^1_\m,P^1_\n]=0\ ,$$
$$\{Q^2_\a,Q^2_\b\} =\G^\m_{\ab} P^2_\m,[P^2_\m,Q^2_\a]
=[P^2_\m,P^2_\n]=0\eqno(2.1)$$
(different signs in the r.h.s. of anticommutators of spinor charges
are chosen for the reason to be clear later).
The odd generators are assumed to be Majorana--Weyl spinors with 16 independent
components so $\Gamma^\mu_{\alpha\beta}$ in (2.1) should actually include the
corresponding projectors.
It will be more convenient for us to impose the Weyl condition on the Grassman coordinates associated
with $Q^1_\alpha,Q^2_\alpha$ rather than
to explicitly incorporate it into the structure relations
(2.1).
One may choose chiralities of $Q^1_\alpha, Q^2_\alpha$ in an
arbitrary way:
the case of the same chiralities corresponds to the IIB type
GS superstring while opposite chiralities lead to the IIA type.
The $D=10$ Lorentz group is treated as an
outer automorphism group of (2.1)
acting on the vector $\mu,\nu,\lambda,\dots$ and spinor
$\alpha,\beta,\gamma,\dots$ indices.
It should be emphasized that (2.1) possesses neither
$SO(2)$ nor $SO(1,1)$ groups of automorphisms;
this becomes clear after
rewriting (2.1) as
$$\{Q^i_\a,Q^j_\b\}=-\G^\m_{\ab}(P^+_\m\s^{ij}_3 +P^-_\m\d^{ij})\ ,$$
$$[P^\pm_\m,Q^i_\a]=0,\
\s_3=\left({1\atop 0}{0\atop -1}\right),\
P^\pm_\m ={1 \over 2}(P^1_\m \pm P^2_\m)\eqno(2.2)$$
For future use, we note that the reductions
$P^+_\mu = 0$ or $P^-_\mu = 0$
in (2.2) lead, respectively, to the two kinds of $N=2$ $D=10$ superslgebras,
the standard one with the $SO(2)$ automorphism
group and the ``noncompact'' one with the $SO(1,1)$ automorphisms.
The presence of the additional vector generator $P^+_\mu$ in (2.2) will turn out
crucial for an algorithmic construction of the WZ term:
the latter will be shown to essentially involve the Cartan form associated with this generator.

B. The target space of our sigma model is identified with the coset
$G_1\otimes G_2/G^+$ where $G^+$ is the abelian translation subgroup
generated by $P^+_\mu$.

This space actually coincides with ordinary $N=2$ $D=10$ superspace.
Indeed, let us parametrize the elements of $G_1,G_2$ as
$$U_j=\exp\left\{i\left({1 \over 2} x^{j\m}P^j_\m +\t^{j\a}Q^j_\a\right)
\right\},\ j=1,2\eqno(2.3)$$
The left action of $G_1,G_2$ on these elements induces the supersymmetry
transformations in the two $N=1$ $D=10$ superspaces $ \{x^{1\m},\t^{1\a}\}$ and $\{x^{2\m},\t^{2\a}\}$
$$x^{1\m'}=x^{1\m}+i\v^{1\a}\G^\m_{\ab}\t^{1\b},\
\t^{j\b'}=\t^{j\b}+\v^{j\b}\ ,$$
$$x^{2\m'}=x^{2\m}-i\v^{2\a}\G^\m_{\ab}\t^{2\b}\ .\eqno(2.4)$$
Considering the product $U_1U_2$ and factoring out the exponent with the $G_+$
generator $P^+_\mu$ one represents elements of $G_1\otimes G_2/G_+$ as
$$\{G_1\otimes G_2/G_+\}=\exp\left\{ix^\m P^-_\m+i\sum_j
\t^{j\a}Q^j_\a\right\}\eqno(2.5)$$
with $$ x^\m =x^{1\m}-x^{2\m}\ . $$
 {}From (2.4) it follows that $G_1\otimes G_2$ acts on the set
$\{ x^\mu,\theta^{1\alpha},\theta^{2\alpha}\}$
in precisely the same way as $N=2\  D=10 $ supergroup
$$x^{\m'}=x^\m +\sum_j\v^{j\a}\G^\m_{\ab}\t^{j\b},\ \t^{j\a'}=\t^{j\a}+\v^{j\a}\eqno(2.6)$$

Another way to realize that $G_1\otimes G_2$ is
undistinguishable from the $N=2$ Poincar\'e
supergroup when applied to
$\{x^\mu,\theta^{j\alpha}\}$ is to
take notice of the fact that $P^+_\mu$ is zero on this
coordinate set and so (2.2)
actually reduces on the latter to $N=2$ supertranslation algebra.
The difference of (2.2) from $N=2$ superalgebra
actually manifests itself only in the existence of an
inhomogeneously transforming Cartan's form on
$\{x^\mu,\theta^{j\alpha}\}$ associated with
$P^+_\mu$ (this is a
$P^+_\mu$ connection).

The construction of a 2D nonlinear sigma model
based on these two assumptions follows the
standard routine.
The building blocks are the left--invariant one--forms on the supergroups
$G_1$ and $G_2$
$$u^{-1}_j(\xi)\p_a U_j(\xi)=\o^{j\m}_aP^j_\m +\o^{j\a}_a Q^j_\a=$$
$$=\left[{i\over 2} (\p_a x^{j\m} -i(-1)^j\p_a\t^j\G^\m\t^j)\right]
P^j_\m +[i\p_a\t^{j\a}]Q^j_\a\ ,\eqno(2.7)$$
where $x^{j\mu}$ and $\theta^{j\alpha}$ are treated as the scalar fields on the
world sheet
$\xi = (\xi^0,\xi^1)$.
The Cartan--forms describing the coset
$G_1\otimes G_2/G_+$ (2.5) are composed of
these objects in an evident way
$$\o^\m_a = \o^{1\m}_a-\o^{2\m}_a={i\over 2}
\left(\p_ax^\m +i
\sum^2_{j=1}\p_a\t^j\G^\m\t^j\right)\ ,$$
$$\o^{j\a}_a=i\p_a\t^{j\a}\ ,$$
$$\O^\m_a=-{1 \over 2}\sum_{j,k}\p_a\t^j\G^\m(\s_3)^{jk}\t^k\eqno(2.8)$$
and are introduced by the generic nonlinear--realization formula  $(x^{1\mu} +x^{2\mu} = 0)$
$$ \o_a\equiv U^{-1}_2(\xi)U^{-1}_1(\xi)\p_a[U_1(\xi)U_2(\xi)]=
\o^\m_aP^-_\m+\sum^2_{j=1}
\o^{j\a}_aQ^j_\a+\O^\m_aP^+_\m\eqno(2.9)$$
Here the Cartan forms
$\omega^\mu_a d\xi^a,\
\omega^{j\alpha}_a d\xi^a$
associated with the coset generators
$P^-_\mu,\
Q^j_\alpha$ are covariant differentials of the
coset fields
$x^\mu(\xi),\
\theta^{j\alpha}(\xi)$.
The one--form $\Omega^\mu_a d\xi^a$ is the connection on the stability
subgroup $G_+$.

It is convenient to single out the coset
$G_1\otimes G_2/G_+$ from the product
$G_1\otimes G_2$ by imposing invariance under the right gauge $G_+$ transformations
$$U_1(\xi)\to U_1(\xi)\exp\{y^\m(\xi)P^1_\m\}\ ,$$
$$U_2(\xi)\to U_2(\xi)\exp\{y^\m(\xi)P^2_\m\}$$
$$\o^{j\m}_a\to \o^{j\m}_a+\p_ay^\m(\xi),\ \o^{j\a}_a\to \o^{j\a}_a\eqno(2.10)$$
Any action invariant under these
 gauge transformations actually involves
only the coset space coordinates
$\{x^\mu,\theta^{j\alpha}\}$.
The Cartan forms (2.8), (2.9) correspond to a particular fixing of this gauge freedom, so as
$$x^{1\mu} =-x^{2\mu} =\half x^\mu$$
The useful object is
$$U(\xi) =U_1(\xi) U^{-1}_2(\xi)=$$
$$=\exp\left\{i\half x^{1\m}P_\m +i\t^{1\a}Q_\a\right\}\exp\left\{-i\half
x^{2\m}P_\m-\t^{2\a}Q_\a\right\}\eqno(2.11)$$
where we have substituted
$P^1_\mu =P^2_\mu = P_\mu$,
$Q^{1\mu} = iQ^{2\alpha} = Q^\alpha$
(formally, $-iQ^\alpha$,
$P^\mu$ satisfy the same superalgebra as $Q^{2\alpha}$,
$P^2_\mu$).
This supergroup element is invariant
under the gauge transformations (2.10) and so lives
on the coset $G_1\otimes G_2/G_+$.
The supergroups $G_1$ and $G_2$
act as the left and right multiplications of $U(\xi)$
$$U'(\xi)=g_1U(\xi)g^{-1}_2,\ g_1\in G_1,\ g_2\in G_2\eqno(2.12)$$
This matrix field is an analog of the standard principal chiral fields,
the only difference is that the latter describes symmetric
cosets of the type
$G_L\otimes G_R/G_{diag}(G_L\sim G_R$
can be arbitrary groups or supergroups) and is invariant
under right gauge $G_{diag}$ transformations
while in the case at hand we deal with the
nonsymmetric coset.
Correspondingly, from $U(\xi)$ one may construct the ``left'' and ``right'' linearly transforming
currents
$$J^L_a(\xi) =U\partial_a U^{-1}=-U_1[\o^\m_aP_\m+(\o^{1\a}_a+i\o^{2\a}_a)Q_\a]U^{-1}_1\ ,$$
$$J^R_a(\xi)=U^{-1}\p_a U=
U_2[\o^\m_a P_\m+(\o^{1\a}_a+i\o^{2\a}_a)Q_a]U^{-1}_2\ ,$$
$$J^-_a=J^{L-}_a=J^{R-}_a=U_1\o^\m_aP_\m U^{-1}_1
=U_2\o^\m_aP_\m U^{-1}_2=\o^\m_aP_\m\ ,$$
$$J^{L'}_a=g_1J^L_ag^{-1}_1\  j^{R'}_a=g_2J^R_ag^{-1}_2\ .\eqno(2.13a)$$
However, in the present case, besides $J^L_a$, $J^R_a$, one may define two more objects
$$J^{Lj}_a=U_1\o^{j\a}_a Q_\a U^{-1}_1\ , J^{Rj}_a=U_2\o^{j\a}_a
Q_\a U^{-1}_2\eqno(2.13b)$$
transforming in the same way as $J^L_a$, $J^R_a$.
This is because our coset space is nonsymmetric and, as a consequence,
there exist on it three independent homogeneously transforming
Cartan's forms
$\omega^\mu_a d\xi^a$,
$\omega^{1\alpha}_a d\xi^a$,
$\omega^{2\alpha}_a d\xi^a$ and each can be used to construct a
current--like quantity.

\subsection{The $N=2$ GS action as a WZW action}
The generic form of the action of WZW sigma associated with the coset
$G_1\otimes G_2/G_+$ is as follows \cite{[9],[10]}
$$A=\int_{\p V}d^2\xi\sqrt{-g}g^{ab}\langle \o_a \o_b\rangle_I+
\half \int_V d^3\xi\v^{ABC}\langle\o_A[\o_B,\o_C]\rangle _{II}\eqno(2.14)$$
Here $\omega_a d\xi^a,\
\omega_A d\xi^A(A=1,2,3)$ are Cartan's forms defined by Eq.(2.9),
$\partial V$ is a two--dimensional boundary of the three--dimensional
region $V$ and $g^{ab}(\xi)$ is a metric on $\partial V$.
The symbols $\langle\dots\rangle_{I,II}$ stand for a kind of the metric on
superalgebras (2.1).
As is well--known, it is impossible to define the cyclic operation $Str$ for the supertranslation algebras (2.1) and, respectively,
an invariant non--degenerate metric.
So, $\langle\dots\rangle_{I,II}$ in (2.14) should be
regarded rather as some general bilinear forms of the Cartan forms
$\omega^\mu_a,\omega^{1\alpha}_a,\omega^{2\alpha}_a,\Omega^\mu_a$ with the
coefficients unspecified for the moment.
These coefficients are almost uniquely (up to two parameters)
fixed by the following four natural principles

1) $D=10$ Lorentz invariance and rigid $G_1,G_2$ supersymmetry;

2) Gauge (local) right $G_+$ invariance;

3) The closeness of the WZ three--form

$$\Omega_3 =\langle\omega\wedge\omega\wedge
\omega\rangle_{II},\ \delta\Omega_3 = d\Omega_2$$

where $\Omega_2$ is a two--form.

4) The absence of the fermionic kinetic terms having the second order in $\partial/\partial\xi^0$.

These principles specify the ``metrics''
$\langle\dots\rangle_{I,II}$ up to two constants $\ell_I,\ell_{II}$
$$\lan P^-_\m P^-_\n\ran_I=\ell_I\eta_{\mn},\lan P^-_\m,P^+_\n\ran_{II}=\ell_{II}\eta_{\mn}\ ,\eqno(2.15)$$
all the remaining ``averages'' of bilinear forms in the generators being zero.
Correspondingly, the action takes the form
$$A=\ell_I\int_{\p V} d^2\xi \sqrt{-g} g^{ab}\o^\m_a\o_{b\m}+$$
$$+\half \ell_{II} \int_V d^3\xi\v^{ABC}\o^\m_A
\o^{j\a}_B\G_{\m,\ab}(\sigma_3)^{ij}\o^{i\b}_C\ ,\eqno(2.16)$$
or
$$A=\ell_I\int_{\p V} d^2\xi \sqrt{-g} g^{ab}\o^\m_a\o_{b\m}+$$
$$+\ell_{II} \int d^3\xi \v^{ABC}\o^\m_A\p_B\O_{C\m}\eqno(2.16')$$
where in achieving two equivalent forms of the WZ term we have
made use of the Maurer--Cartan equations for $\omega_A$
to rewrite the product of the spinor 1--forms
as $\partial_{[B}\Omega^\mu_{C]}$.
Note a natural appearance of the matrix
$(\sigma_3)^{ij}$ in the WZ term (2.16)
as the structure constants standing before the generator
$P^+_n$ in the superalgebra (2.2)
we have started with.
This matrix has come out as the result of the evaluation
 of the commutator
$[\omega_B,\omega_C]$ in (2.14).
At the same time, in the standard view on $N=2$ GS superstring, it is introduced to some extent,
``by hand'' \cite{[7],[8]} because it is not present in
$N=2$ $D=10$ supertranslation algebra.
The breaking of $SO(2)$ automorphism symmetry in the WZ term gets now an
explanation as connected with the fact that the superalgebra (2.2)
possesses no internal automorphisms at all
(neither $SO(2)$ nor $SO(1,1)$).
A subtle point is, of course, a specific choice of the ``metrics''
$\langle\dots\rangle_{I,II}$ in (2.15)
which cannot be immediately related to the inner structure
 of superalgebra (2.2).
An analogy with the WZ terms of ordinary sigma models
becomes even more direct in the case of non--abelian superstrings constructed along similar lines in the next section.
For nonabelian analogs of superalgebra (2.2) one is able to define the cyclic operation
$Str$ and, respectively, a nondegenerate metric, so the symbols
$\langle\dots\rangle_{I,II}$ become well defined in this case.
Upon contraction to the flat superalgebra (2.2), one finds that the
limiting action involves just the ``metrics'' (2.15). Thus
the choice (2.15) can in fact be justified by resorting to the nondegenerate
case of nonabelian superstrings.

We wish to point out that the second translation generator
$P^+_\mu$ manifests itself in the present context only
in giving rise to an additional Cartan form $\Omega^\mu$ which underlies
the $N=2$ GS WZ term.
In this aspect, $P^+_\mu$ resembles the central charge operators in the geometric
formulations of $N\geq 2$ 4D supergravities.
These operators serve to render
a geometric meaning to some members of the relevant gauge multiplets
(e.g. to the graviphoton in $N=2$ Einstein supergravity),
however, in their own right, do not generate any symmetry of the latter.
Note that the nonabelian analogs of $P^+_\mu$ possess a
nontrivial action on the target superspace coordinates
(see next Section).

Let us come back to the discussion of (2.16).
Rewriting the WZ term as an integral over
$\partial V$ with the help of the important
identity for Dirac $D=10$ $\gamma$ matrices

$$\Gamma^\mu_{\alpha,\beta}\Gamma_{\mu,\gamma\delta}
 + ({\rm cyclic}\  (\alpha,\beta,\gamma)) = 0\eqno(2.17)$$
we obtain the GS covariant ``preaction'' containing
two free parameters $(\v^{01} =-\v^{10} = 1)$
$$A=-{\ell_I\over 4}\int_{\p V}d^2\xi
\left\{\sqrt{-g}g^{ab}\left(\p_ax^\m+i\sum^2_{j=1}
\p_a\t^j\G^\m\t^j\right)\left(\p_bx_\m+
i\sum^2_{j=1}
\p_b\t^j\G_\m\t^j\right)-\right.$$
$$\left.-{\ell_{II}\over \ell_I} \v^{ab} \left(i\p_ax^\m-\half\sum^2_{j=1}
\p_a\t^j\G^\m\t^j\right)\left(\sum^2_{k,j=1}\p_b\t^k\s^{kj}_3
\G_\m\t^j\right)\right\}\eqno(2.18)$$

The genuine GS action possessing local $\kappa$--supersymmetry
arises at
$\ell_{II}/\ell_I = 2$ \cite{[6]}.
The $\kappa$--symmetry transformations are given by
$$\tilde\o^{1\m} +\tilde\o^{2\m}=0$$
$$\tilde\o^{1\a}=P^{ab}_+ (\o^1_{b\m}-\o^2_{b\m})
\tilde\G^{\m,\ab}\kappa^1_{a\beta}(\xi)$$
$$\tilde\o^{2\a}=P^{ab}_-(\o^1_{b\m}-\o^2_{b\m})
\tilde\G^{\m,\ab}\kappa^2_{a\beta}(\xi)$$
$$\d(\sqrt{-g}g^{ab})=-2(P^{da}_+P^{cb}_+\o^{1\a}_d
 \kappa^1_{c\alpha }+
P^{da}_-P^{cb}_-\o^{2\a}_d\kappa^2_{c\a})\eqno(2.19)$$
where $\kappa^j_{\alpha\beta}$ are transformation parameters,
$\tilde\Gamma^\mu = C^{-1}\gamma^\mu$, $P^{ab}_\pm$ are
2D light--cone projectors
and we have introduced the left--invariant variations
$$(U_j)^{-1}\delta U_j =\tilde\omega^{j\mu}P^j_\mu+\tilde\omega^{j\alpha}Q^j_\alpha$$

\subsection{A zero--curvature representation for the GS superstring equations}

In terms of Cartan's forms the GS superstring equations of motion are written as
$$\p_a[P^{ab}_+\o^\m_b-{\ell_{II}\over \ell_I}
\v^{ab}\o^{1\m}_b]=0\ ,\eqno(2.20)$$
$$\p_a[P^{ab}_-\o^\m_b-{\ell_{II}\over \ell_I} \v^{ab}\o^{2\m}_b]=0\ ,$$
$$P^{ab}_+\o_{b\m}\G^\m_{\ab}\o^{1\b}_a=0\ ,$$
$$P^{ab}_-\o_{b\m}\G^\m_{\ab}\o^{2\b}_a=0\eqno(2.21)$$
where $P^{ab}_\pm =\sqrt{-g}g^{ab}\pm
{\ell_{II}\over 2\ell_I}\varepsilon^{ab}$.
We have to add to (2.20) and (2.21) the Maurer--Cartan
equations and the equation
$$\o^\m_a\o_{b\m}-\half g_{ab}g^{cd}
\o^\m_c \o_{d\m}=0\eqno(2.22)$$
which is obtained by varying $g^{ab}$ in (2.16).
The set of the basic GS superstring Eqs.(2.20), (2.21) admits a reformulation as the conservation laws for
the appropriate $G_1$ and $G_2$ currents
(with the Maurer--Cartan equations taken into account)
$$\p_a[P^{ab}_+\tilde J^1_b-{\ell_{II}\over \ell_I}
\v^{ab}\tilde J^{L1}_b]=0\ ,$$
$$\p_a[P^{ab}_-\tilde J^2_b -{\ell_{II}\over \ell_I}
\v^{ab}\tilde J^{R2}_b]=0\ ,\eqno(2.23)$$
where the quantities in the square brackets are
$$\tilde J^j_b =\o^\m_b P_\m +\o^\m_b(\t^j\g_\m Q)$$
$$\tilde J^{L1}_b =\o^{1\m}_bP_\m +{1\over 3}
(\t^1\G_\m\p_b\t^1)(\t^1\g^\m Q)$$
$$\tilde J^{R2}_b = \o^{2\m}_b P_\m -
{1\over 3}(\t^2\G_\m\p_b\t^2)(\t^2\g^\m Q) $$

Surprisingly, the GS Eqs.(2.20), (2.21) admit a zero
curvature representation, precisely at the same value of
$\ell_{II}/\ell_I = 2$ which is selected by $\kappa$--invariance.
At this specific value of $\ell_{II}/\ell_I$
one may rewrite these equations as the
integrability conditions
$$[L^1_a,L^1_b]=[L^2_a,L^2_b]=0\ ,\eqno(2.24)$$
$$L^1_a=\p_a-2\l^2\v_{ab}P^{bc}_+\o^\m_cR^1_\m+2
\l i\o^{1\a}_aS^1_\a\ ,$$
$$L^2_a=\p_a+2\l^2\v_{ab}P^{bc}_-\o^\m_cR^2_\m+2\l i
\o^{2\a}_aS^2_\a\ ,\eqno(2.25)$$
where $\lambda$ is a spectral parameter and
$(R^j_\mu,S^j_\alpha)$ constitute two
mutually commuting superalgebras \cite{[9]}
$$[R^j_\m,R^j_\n]=R^j_{[\mn]}\ ,\eqno(2.26a)$$
$$[R^j_\m,S^j_\a]=\G_{\m,\ab}\tilde S^{j\b}\ ,\eqno(2.26b)$$
$$\{S^j_\a, S^j_\b\} =-\G^\m_{\ab}R^j_\m\ .\eqno(2.26c)$$
Note that the generator $\tilde S^{j\beta}$ cannot be put proportional to
$S^{j\beta}$ without contradiction between
Eqs. (2.26b), (2.26c) and the assumption that
$R^j_{[\mu\nu]}\not= t^\lambda_{\mu\nu}R^j_\lambda$.
Actually only the relations (2.26b) and (2.26c) are of immediate relevance for deducing Eqs.(2.20),
(2.21) from (2.24), (2.25).
The form of the relation (2.26a) is not fixed
(nonzero $R^j_{[\mu\nu]}$ seem to be of need only while
constructing an infinite number of the conserved
currents associated with the Lax pair (2.24)).
Also, the newly introduced generator
$\tilde S^{j\beta}$
is not obliged to commute
with the old ones and with
itself.
So, the complete zero--curvature representation superalgebra can be in principle infinite--dimensional.
The minimal possibility which yet allows one to establish the
equivalence of Eqs.(2.20), (2.21) with (2.24) and (2.25) is to put
$R^j_{[\mu\nu]}=0$,
$[\tilde S, R_\mu] =\{\tilde S,\tilde S\} = 0$
in (2.24), (2.26). The corresponding superaglebra is known now as Green's superalgebra \cite{[12]}.
Thus, the $N=2$  GS superstring equations admit a minimal zero--curvature representation on the two
mutually commuting Green's superalgebras.

It is interesting to mention that in the language of the zero--curvature representation the local
$\kappa$--symmetry of the GS equations reveals itself as a kind
of gauge transformations of the operators
$L^1_a$,
$L^2_a$
preserving the representation (2.24)
$$L^j_a\to G_j(\xi)L^j_a G^{-1}_j(\xi)\eqno(2.27)$$
where, in the infinitesimal form,
$$G_1(\xi)=1-2i\l \tilde\o^{1\a}S^1_\a$$
$$G_2(\xi)=1-2i\l\tilde\o^{2\a}S^2_\a\eqno(2.28)$$
and variations $\tilde\o^{j\a}$ were defined
in Eqs.(2.19)
(when checking invariance of (2.20), (2.21) under (2.27), (2.28), one has to take
into account the equations of motion).

\subsection{Summary and comments}
To summarize, the sigma--model interpretation of the
covariant action of $N=2$ GS superstring is most naturally achieved provided one
starts with the product of two
$N=1$ D=10 Poincar\'e
supertranslation groups $G_1\otimes G_2$ and
constructs the WZW sigma model on the coset space
$G_1\otimes G_2/G_+$ where $G_+$ is the diagonal translation subgroup.

It is natural to ask how the present construction could be extended to include the supergroups
different from the $N=1$ supertranslation ones, in particular those with a nonabelian even part,
and how all this matches with $N=2$ GS superstring moving in an arbitrary $N=2$ $ D=10$
supergravity background. The first question was partly addressed in our paper \cite{[10]} where we
considered the superstring--like sigma models associated with the more general superalgebras of the
type (2.10) which were obtained by contraction from arbitrary Lie superalgebras. In this way we
have arrived at the actions of the kind (2.17) in which, however, the structure constants
$\Gamma^\mu_{\alpha\beta}$ do not in general
coincide with $\gamma^\mu C$.
These models still reveal the property of classical integrability,
but the local $\kappa$--symmetry is present only under some
severe restrictions on the constants
$\Gamma^\mu_{\alpha\beta}$.

In \cite{[9],[10]} we have regarded $G_1$ and $G_2$ to be ``abelian'', i.e. possessing the
superalgebras of the type (2.1), in which the even generators commute with each other and with the
odd ones. In the next section we shall study the most general situation, with the above
restrictions on
$G_1, G_2$ removed.
The construction outlined here will be
shown to work equally well also in this general case.
It yields a new wide class of self--consistent
WZW sigma models admitting an
interpretation in terms of the superstring moving in a nonabelian supergroup background.
\setcounter{equation}{0}

\section{THE GREEN--SCHWARZ TYPE SUPERSTRING ON THE LIE SUPERGROUP COSET SPACE}

Our aim here is to generalize the construction expounded in Sec.2
to the case of an arbitrary Lie supergroup with nondegenerate
metric. The resulting sigma model will be interpreted as a theory of
superstring moving on a curved coset supermanifold.

\subsection{Supergroup preliminaries}

Let $G_1(d\vert D)$ be a Lie supergroup with the even and odd
generators $R^1_\mu(\mu,\nu,\lambda,\dots = 1,2,\dots d)$ and
$S^1_\alpha(\alpha,\beta,
\gamma,\dots = 1,2,\dots,D)$
which obey the following (anti)commutation relations
$$[R^1_\m,R^1_\n]=t^\l_{\mn}R^1_\l\ , \{S^1_\a,S^1_\b\}=
-\G^\m_{\ab}R^1_\m, [R^1_\m,S^1_\a]=C^\b_{\m\a}S^1_\b\ ,
\eqno(3.1)$$
$\Gamma^\mu_{\alpha\beta},C^\beta_{\mu\alpha},t^\lambda_{\mu\nu}$ being the structure constants.
The only restriction we impose on (3.1) is the existence of a nondegenerate metric
$$\eta_{\mn}=\eta_{\nu\mu}=-Str(R^1_\m R^1_\n),\
X_{\ab}=-X_{\b a}
=Str(S^1_\a S^1_\b)\ , Str(R^1_\m S^1_\a)=0 $$
Then the evident identity
$$Str(\{S^1_\a, S^1_\b\}R^1_\m)= Str(S^1_\a[S^1_\b  ,R^1_\m]) $$
yields [10]
$$\G^\m_{\ab}\eta_{\mn}=C^\g_{\mu\b}X_{\g\a} $$
or
$$C^\a_{\m\b} =\eta_{\mn}\G^\n_{\b\g}X^{\g\a}\ (X_{\a\g}X^{\g\b}=\d^\b_\a) $$
 {}From the identity
$$Str(\{S^1_\a, S^1_\b\}\{S^1_\g,S^1_\d\}+
({\rm cyclic}\  \a,\b,\g))=0 $$
one deduces the important relation (compare with Eq.(2.17))
$$[\G^\m_{\ab}\G^\n_{\g\d}+({\rm cyclic}\ \a,\b,\g)]
\eta_{\mn} = 0\eqno(3.2)$$
 {}From the Jacobi identities there also follow certain quadratic
 relations between
the structure constants. These can be easily derived, so we do not give them explicitly (see, e.g.
\cite{[10]}). The described class of  superalgebras is very wide: it includes all the semisimple Lie
superalgebras and also some non-semisimple ones, e.g. those of the type considered by Green
\cite{[12]}. Given the superalgebra (3.1) one may construct a new superalgebra by the formal
substitution
$S^1_\alpha\to iS^2_\alpha$ in (3.1)
$(R^1_\mu \to R^2_\mu)$
$$[R^2_{\m}R^2_\n]=t^\l_{\mn}R^2_\l,
\{S^2_\a,S^2_\b\} =+\G^\m_{\ab}R^2_\m,
[R^2_\m, S^2_\a]=C^\b_{\m\a}S^2_\b\eqno(3.3)$$ This superalgebra is called dual in Cartan's sense
to (3.1) \cite{[11]}\footnote{Dual
(super)algebra can be defined for any (super)algebra admitting a
$Z_2$ grading.
For instance, dual to$SU(2)$ is $SU(1,1)$.}.
Both (3.1) and (3.3) may be regarded as different real forms of the superalgebra
$g^c(d\vert D)$ of the complex supergroup
$G^c(d\vert D)$ generated by the double set of generators
$(R^1_\mu,iR^1_\nu),\
(S^1_\alpha, iS^1_\beta)$.
It is easy to check that the pairs
$(R^1_\mu, S^1_\alpha)$ and
$(R^1_\mu,iS^1_\beta)$
constitute in $g^c$ two subsuperalgebras which are isomorphic to (3.1)
and (3.3) and have as a crossover the even subalgebra with the generators
$R^1_\mu$.
In what follows we shall regard (3.1) and (3.3)
as two independent mutually commuting Lie superalgebras.

Generally, (3.1) and (3.3) are not isomorphic to each other.
They become isomorphic in the degenerate limit
$C^\beta_{\mu\alpha} =t^\lambda_{\mu\nu} = 0$
corresponding to the case of flat superspace treated in Sec.2 (in this case, (3.3) takes the form
(3.1) after the redefinition $R^2_\mu\to -R^2_\mu)$\footnote{The superalgebras (3.1) and (3.3) can
also be isomorphic if $t^\lambda_{\mu\nu} = 0$, $C^\beta_{\mu\alpha}\not= 0$. An example of that
sort is given by Green's superalgebra \cite{[12]}.}

The basic motivation for considering the dual superalgebra (3.3) in parallel with (3.1) comes from
the desire to have a correct generalization of the flat superspace construction of Sec.1 to curved
case. As it will become clear later, this is possible only provided one construct the WZW sigma
model on the product $G=G_1\otimes G_2$ where the supergroups $G_1$ and $G_2$ correspond to dual
superalgebras (3.1) and (3.3) and so are dual to each other. Only under this choice, in the flat
superspace limit one recovers the standard GS superstring type model (see Subsec.2.4). {}From the
mathematical point of view,
$G=G_1\otimes G_2$ is distinguished in that it is self--dual in Cartan's sense.

\subsection{Nonlinear realization of $G_1\otimes G_2$}

In constructing the $G_1\otimes G_2$ WZW sigma model we shall closely follow the consideration in Sec.2.
The elements of $G_1$ and $G_2$ are
parametrized as
$$U_1 =\exp\left\{{i\over 2} x^{1\m}R^1_\mu  +
i\t^{1\a}S^1_\a\right\}$$
$$U_2 =\exp\left\{{i\over 2} x^{2\mu} R^2_\mu  + i\t^{2\a}S^2_\a
\right\}=
\exp\left\{{i\over 2}x^{2\m}R^2_\m +\t^{2\a}S_\a\right\}\eqno(3.4)$$
$$S_\a=iS^2_\a,\ \{S_\a, S_\b\} =-\G^\m_{\ab}R^2_\m,\
\{S^j_\a,\t^{k\b}\}=0\eqno(3.5)$$
with $\{x^{1\mu},x^{2\mu}\}$ and $\{\theta^{1\alpha},\theta^{2\alpha}\}$ being even and odd real supergroup parameters.
it is seen that $U_1$ and $U_2$ can be viewed as two different restrictions of a
generic element of the complex supergroup $G^c(d\vert D)$
$$U^c =\exp\left\{{i\over 2}(x^\m +iy^\m)R_\m +i(\t^\a +i\eta^\a)
S_\a\right\}$$
$$U^c\to U_1 \ {\rm if}\  y^\m = 0, \ \eta^\a=0$$
$$U^c\to U_2 \ {\rm if}\  y^\m = 0,\ \t^\a=0 $$

Further, $U_1$ and $U_2$ are assumed to be defined on the world sheet $(\xi^0,\xi^1)$
$$U_i = U_i(\xi^0,\xi^1) $$
To have a kind of $N=2$ superspace as the sigma model target space, we should identify the latter with the homogeneous space
$G_1\otimes G_2/G_+$ where $G_+$ is the diagonal even subgroup generated by
$$R^+_\m =\half(R^1_\m +R^2_\m)\eqno(3.6)$$
The elements of the coset $G_1\otimes G_2/G_+$
are represented by the group orbit
$$U_1U_2\exp\{iy^\m(\xi)R^+_\mu\} $$
where $y^\mu(\xi)$ are arbitrary world--sheet fields.
So, the functionals defined on the coset space
$G_1\otimes G_2/G_+$ are singled out from
those on the whole supergroup $G_1\otimes G_2$ by requiring them to be invariant under the right gauge $G_+$ transformations
$$U_j\to U_j H_j(\xi),\
H_j(\xi) =\exp({i\over 2} y^\m(\xi)R^j_\m)\ , j = 1,2\ .\eqno(3.7)$$
One may fix the gauge so that $G_1\otimes G_2/G_+$ is
 parametrized by the $N=2$ superspace type coordinates
$$G_1\otimes G_2/G_+ =\left\{x^\mu =\half(x^{1\m}-x^{2\m}),
\t^{1\a},\t^{2\b}\right\}\eqno(3.8)$$
associated with the coset generators $R^-_\mu =\half(R^1_\mu-R^2_\mu)$,
$S^1_\alpha,S^2_\beta$.
Note that the superspace $G_1\otimes G_2/G_+$ (like its
flat prototype) is not symmetric: (anti)commutators of the
coset generators contain in their r.h.s.
the generators of the same sort, alongside with those of the
stability subgroup $G_+$. The supergroups $G_1$ and $G_2$
are realized on the elements of the coset $G_1\otimes G_2/G_+$
by left shifts which are accompanied, in the particular gauge (3.8),
by induced right $G_+$
transforamtions. As distinct from the flat abelian case, the subgroup $G_+$ has a nontrivial action on coordinates (3.8).
Therefore, the product $G_1\otimes G_2$ is by no means reduced to anything like $N=2$ supergroup while realized
on these coordinates.

The next step is the construction of Cartan's forms.
Like in Sec.2 we first define the left--invariant one--forms on supergroups $G_1$ and $G_2$ separately
$$U^{-1}_j\p_aU_jd\xi^a=\o^j=\o^j_a d\xi^a =
(\o^{j\m}_aR^j_\m +\o^{j\a}_a S^j_\a) d\xi^a\eqno(3.9)$$
(no summation over $j=1,2$!).
They are invariant under global left $G_1$ and $G_2$
 tranformations
$$U_j \to \exp(i\v^{j\a}S^j_\a)U_j$$
$$U_j\to\exp(ia^{j\mu}R^j_\mu)U_j\eqno(3.10)$$
which are a nonabelian generalization of $N=1$ supertranslations
and ordinary translations.
Under the gauge transformations (3.7) $\omega^j$ transform as gauge connections
$$\o^j\to H^{-1}_j\o^jH_j+H^{-1}_j dH_j\eqno(3.11)$$

In what follows, it will be convenient to relate these
one--forms to the same set of generators using the fact
that $(R^1_\mu, S^1_\alpha)$ and $(R^2_\mu, iS^2_\alpha)$
obey the same structure relations. One has
$$\o^1 =\o^{1\mu} R_\mu +\o^{1\a}S_\a,\
\o^2=\o^{2\mu}R_\mu -i\o^{2\a}S_\a\eqno(3.12)$$
$$\o^{i\m}R_\m\to H^{-1}\o^{i\m}R_\m H+H^{-1}dH$$
$$\o^{i\a}S_\a \to H^{-1}\o^{i\a}S_\a H,\ H = \exp\left(i{y^\mu(\xi)\over 2} R_\m\right)\eqno(3.13)$$
Note that the structure of $U_1$ and $U_2$ implies that $\omega^1,\omega^2$ are related via simple changes of supergroup
coordinates
$$\o^{1\m}\lrarrow \o^{2\m},\quad x^{1\m}\lrarrow x^{2\m}$$
$$\o^{1\a}\lrarrow -i\o^{2\a},\quad
\t^{1\a}\lrarrow -i\t^{2\a}\eqno(3.14)$$
For future use, we write down the Maurer--Cartan equations which follow from the definitions (3.9) of $\omega^j$
$$d\o^j +\o^j\wedge \o^j=0\eqno(3.15)$$
or, in terms of the coefficients of one--forms,
$$\p_a\o^{1\m}_b-\p_b\o^{1\m}_a+\o^{1\r}_a\o^{1\n}_b
t^\m_{\r\n}+
\o^{1\a}_a\o^{1\b}_b\G^\m_{\a\b}=0\ ,$$
$$\p_a\o^{1\a}_b-\p_b\o^{1\a}_a+(\o^{1\m}_a\o^{1\b}_b-
\o^{1\m}_b\o^{1\b}_a)C^\a_{\m\b}=0\eqno(3.15a)$$
$$\p_a\o^{2\m}_b-\p_b\o^{2\m}_a+\o^{2\r}_a\o^{2\n}_bt^\m_{\r\n}
-\o^{2\a}_a\o^{2\b}_b\G^\m_{\ab}=0\ ,$$
$$\p_a\o^{2\a}_b-\p_b\o^{2\a}_a+(\o^{2\m}_a
\o^{2\b}_b-\o^{2\m}_b \o^{2\b}_a)
C^\a_{\m\b}=0\ .\eqno(3.15b)$$
The one--forms $\omega^j$ are the convenient building blocks out of which one may construct the Cartan forms on the coset
$G_1\otimes G_2/G_+$ and the current one--forms transforming according to the adjoint representaion of $G_1$ and $G_2$.
The Cartan forms having the homogeneous transformation laws with respect to the right gauge $G_+$ group
$$\o^{-\m}R_\m=(\o^{1\m}-\o^{2\m})R_\m,\ \o^{j\a}S_\a$$
$$\o^{-\m}R_\m\to H\o^{-\m}H^{-1},\ \o^{j\a}S_\a\to H
\o^{j\a}S_\a H^{-1}\eqno(3.16)$$
are interpreted as the covariant differentials
of the coset coordinates $(x^\mu,\theta^{1\alpha},\theta^{2d})$
while the inhomogeneously transforming quantity
$$\o^{+\m}R_\m =\half(\o^{1\m}+\o^{2\m})R_\m$$
$$\o^{+\m}R_\m\to H^{-1}\o^{+\m}R_\m H+H^{-1}dH\eqno(3.17)$$
defines the $G_+$ connection.

Further, let us also relate $U_1$ and $U_2$ (2.4) to the same set
 of generators
$R_\mu,S_\alpha$ and define a $G^c$ matrix field
$$U(\xi)=U_1(\xi)(U_2(\xi))^{-1}=\exp({i\over 2}x^{1\m}R_\m+i\t^{1\a}S_
\a)\exp(-{i\over 2}x^{2\m}R_\m-\t^{2\a}S_\a)\eqno(3.18)$$
This field is manifestly invariant under $G_+$ gauge transformations (3.7) (related to the same generators) and so is
 defined on the coset space $G_1\otimes G_2/G_+$ (it actually depends on
$(x^\mu,\theta^{1\alpha},\theta^{2\alpha}))$.
The rigid supergroups $G_1$ and $G_2$ act on $U$ as the left and right shifts
$$U'=g_1U(g_2)^{-1},\
g_1\in G_1,\
g_2\in G_2\eqno(3.19)$$
Thus, $U$ is an analog of the chiral field of the principal chiral field sigma models.

The one--forms
$$J^L=UdU^{-1},\ J^R = U^{-1}dU=-U^{-1}J^LU\eqno(3.20)$$
transform linearly, according to the adjoint representations of
$G_1$ and $G_2$
$$J^L\to g_1J^L(g_1)^{-1},\
J^R\to g_2 J^R(g_2)^{-1}\eqno(3.21)$$
These objects (more precisely, their coefficients
$J^L_a,J^R_a$) are similar by the transformation properties to the left
and right currents of sigma models for principal chiral field which correspond to the symmetric cosets
$G\otimes G/G_{diag}$.
Owing to the fact that the coset space $G_1\otimes G_2/G_+$ is nonsymmetric, we have
 more freedom in defining the homogeneously transforming objects
of the type (3.20).
Namely, one may consider the objects
$$J^{L-}=U_1(\o^{-\m}R_\m)U^{-1}_1,\
J^{Li}=U_1(\o^{i\a}S_\a)U^{-1}_1\ ,$$
$$J^{R-}=U_2(\o^{-\m}R_\m)U^{-1}_2, J^{Ri} =U_2(\o^{i\a}S_\a)U^{-1}_2\eqno(3.22)$$
These quantities are manifestly  $H$--invariant and transform according to the
adjoint representations of $G_1$ and $G_2$.
The previously defined currents $J^L$, $J^R$ are specific combinations of (3.22)
$$J^L = -U_1(\o^1-\o^2)U^{-1}_1\equiv -U_1\o^-U^{-1}_1=$$
$$=-U_1[\o^{-\m}R_\mu +\o^{1\a}+i\o^{2\a})S_\a]U^{-1}_1=-J^{L-}-(J^{L1}+iJ^{L2})$$
$$J^R = U_2\o^-U^{-1}_2=J^{R-} +(J^{R1}+iJ^{R2}) \eqno(3.23)$$
Note that $J^L$ and $J^R$, in contrast to the currents
of principal chiral field sigma models,
do not obey simple conservation laws. The conserved currents generating
$G_1$ and $G_2$ symmetries are composed from the basic
quantities (3.22) in a different way (see subsec 3.5).
Nevertheless, $J^L$ and $J^R$ are still useful
objects while constructing the invariant actions.

\subsection{Invariant WZW action}

Now we have all the necessary ingredients to arrange the
 $G_1\otimes G_2$ invariant WZW type actions.
The simplest invariants are those bilinear in the homogeneously transforming
forms
$\omega^{-\mu}_a, \omega^{j\alpha}_a$
(covariant derivatives of the coset coordinates).
The general expression for this part of the Lagrangian can be written as
$${\cal L}_0 =\sqrt{-g} g^{ab} (\xi)[Str(\o^-_a\o^-_b) +
\sum^2_{i,j=1}
\kappa_{ij}Str(\o^{i\a}_aS_\a\o^{j\b}_bS_\b)]\eqno(3.24)$$
with $\kappa_{ij}$ being constants arbitrary for the moment and
$\omega^-_a =\omega^1_a -\omega^2_a$.
In general, this expression is not real and gives rise to the second order
derivative
kinetic terms both for even $x^\mu(\xi)$ and odd $\theta^{1\alpha}(\xi),\
\theta^{2\alpha}(\xi)$ coset space coordinates.
The first difficulty is circumvented by choosing
$$\kappa_{12}+\kappa_{21} =-2i\eqno(3.25)$$

The second peculiarity seems not too fatal
while staying at the pure sigma model level
$(\theta^{1\alpha}(\xi),\theta^{2\alpha}(\xi)$
are world--sheet scalars).
However, keeping in mind going to the abelian GS
superstring limit (see Subsec.3.4), such terms are unwanted as they  could produce the second order equations of  motion for the
physical world--sheet fermions in the light--cone gauge.
So, we leave in (3.24) only the term
 bilinear in $\omega^{-\mu}_a$
$$\kappa_{11} = -1,\ \kappa_{22} = 1\ ,\eqno(3.26)$$
$${\cal L}_0 =-\sqrt{-g}g^{ab}\o^{-\m}_a\o^-_{b\m}\ .
\eqno(3.27)$$
It is worth noting that such an opportunity exists
only because we deal with the nonsymmetric space
$G_1\otimes G_2/G_+$.
Owing to this choice the coset Cartan 1--forms
$\omega^{-\mu},\omega^{j\alpha}$ are assigned to a
reducible representation of the stability subgroup
$G_+$ and, as a result, are separately covariant.
This is a crucial difference from the principal chiral field
sigma models where the underlying coset space
$G\otimes G/G_{diag}$ is symmetric.
There , the Cartan 1--forms valued in the coset
constitute an irreducible representation of the stability subgroup
$G_{diag}$.
So, only the invariants having the form of the first term in (3.25) are admissible in such models.
Note that the WZW model discussed recently by Green \cite{[12]} belongs to this latter class.

Let us now turn to construction of the relevant
WZ term.
Once again, due to the fact that
$\omega^{-\mu}$ and $\omega^{j\alpha}$ are covariant
separately,
one has more possibilities for constructing this term
compared to the $G\otimes G/G_{diag}$ case.
We choose it in the following way
$$WZ=\int_V\O_3 =\int_V d^3\xi \v^{ABC}
Str[\o^-_A\o^-_B\o^-_C+
\sum^2_{i,j=1}
\tilde\kappa_{ij}\p_A(\o^{i\a}_B
S_\a\o^{j\b}_CS_\b)]\eqno(3.28)$$
with constants $\kappa_{ij}$ being arbitrary for the moment.
It is easy to check that the three--form $\Omega_3$
is invariant under global $G_1,G_2$ and gauge $G_+$ transformations
and satisfy the standard closeness condition
$$\d\O_3 =d\O_2\eqno(3.29)$$
where $\Omega_2$ is a two--form.
Actually, these requirements are met by the two pieces in (3.28) separately.
The first piece is an analog of the standard
$G\otimes G/G_{diag}$ WZ term while the second one is new.
The possibility to add it is related to the
property that the one--forms $\omega^{i\alpha}_b d\xi^b$ are covariant in their own right.

Leaving aside the question of how general (3.28) is and anticipating the analysis of Subsec.3.4,
we point out that the ansatz (3.28) is sufficient for ensuring a correct flat
superspace limit of the action.
In fact, the constants $\tilde\kappa_{ij}$
in (3.28) are severely restricted by the reality condition. Using the MC equations (3.15),
it is straightforward to find
$$\v^{ABC} Str(\o^-_A\o^-_B\o^-_C) =-\v^{ABC}\biggl[
\half\o^{-\m}_A\o^{-\n}_B\o^{-\r}_Ct_{\m\n\r}+$$
$$+{3\over 2}(\o^{1\a}_A\o^{1\b}_B-\o^{2\a}_A
\o^{2\b}_B
+2i\o^{1\a}_A\o^{2\b}_B)\o^{-\rho}_C\G_{\r,\ab}\biggr]\ ,\eqno(3.30)$$
$$\sum^2_{i,j=1}\tilde\kappa_{ij}\v^{ABC}\p_A Str(\o^{i\a}_B
S_\a\o^{j\b}_CS_\b)=$$
$$=(\tilde\kappa_{12} -\tilde\kappa_{21})\v^{ABC}(\o^{-\m}_C\o^{1\a}_A\o^{2\b}_B)\G_{\m,\ab}\ .\eqno(3.31)$$
Thus, the imaginary part is absent in (3.28) with the choice
$$\tilde{\kappa}_{12}-\tilde{\kappa}_{21}=3i\eqno(3.32)$$
(as is seen from (3.31), parameters $\tilde{\kappa}_{ii}$ and $\tilde{\kappa}_{zz}$ drop out altogether).

The invariant action obtained as a sum of (3.27) and (3.28)
involves at this step two free parameters $\ell_{I}$ and $\ell_{II}$ (cf. Eq.(2.16))
$$A=-\ell_I\int_{\p V} d^2\xi\sqrt{-g} g^{ab}(\o^{-\m}_a\o^-_{b\m})-$$
$$-\half \ell_{II}\int_V d^3\xi\v^{ABC}
\{\o^{-\m}_A\o^{-\n}_B\o^{-\r}_C t_{\mu\n\r}+
3(\o^{1\a}_A \o^{2\b}_B -\o^{2\a}_A\o^{2\b}_B)
\o^{-\r}_C\G_{\r,\a\b}\}\eqno(3.33)$$
or
$$A=\int_{\p V}d^2\xi\left\{ -\ell_I\sqrt{-g}\ g^{ab}(\omega^{-\mu}_a
\omega^-_{b\mu})-3\ell_{II}\varepsilon^{ab}(\omega^{1\mu}_a
\omega^2_{\b\mu})\right\} +$$
$$+\ell_{II}\int_V d^3\xi\ \v^{ABC}\{\p_A\o^{1\m}_B
\o^1_{c\mu}-\p_A\o^{2\m}_B\o^2_{C\m}+$$
$$+(\o^{1\a}_A\p_B\o^{1\b}_C +\o^{2\a}_A \p_B\o^{2\b}_C)
X_{\ab}\}\eqno(3.33')$$
(the second representation is more convenient when varying $A$).
In order to fix $\ell_{I}$ and $\ell_{II}$, one needs to consider the flat superspace limit of (3.33).

\subsection{The $N=2$ GS type superstrings
as the limiting case of the $G_1\otimes G_2/G_+$ WZW
sigma models}

We accept a natural requirement that (3.33)
goes over to the N=2 GS superstring type action
(2.18) in the flat limit, when
the superalgebras (3.1), (3.3) contract into a sum of N=1
supertranslation algebras
$$\{Q^1_\a, Q^1_\b\}=-\G^\m_{\ab}P^1_\m,[Q^1_\a,P^1_\m]
=[P^1_\m,P^1_\nu]= 0\ ,$$
$$\{Q^2_\a,Q^2_\b\} =\G^\m_{\ab}P^2_\mu,
[Q^2_\a,P^2_\m] =[P^2_\m, P^2_\nu] = 0\eqno(3.34)$$ We call this case ``abelian'' in view of
commutativity of bosonic generators. As was explained in Sec.2, in general
$\Gamma^\mu_{\alpha\beta}$ in (3.34) do not necessarily coincide with Dirac matrices figuring in
the genuine $N=2$ GS action \cite{[10]}. It is also worth recalling that the contracted dual
superalgebras are isomorphic to each other, as opposed to their nonabelian prototypes (3.1), (3.3).

The contraction of (3.1) and (3.3) can be effected by introducing a parameter $c$
$$R_\mu = c^2P_\m,\ S_\a = c Q_\a\ ,\eqno(3.35)$$
$$\{Q_\a,Q_\b\}=-\G^\m_{\ab}P_\m,
[P_\m,Q_\a]={1\over c^2}
C^\b_{\m\a} Q_\b,[P_\m,P_\n]={1\over c^2}
t^\l_{\m\n}P_\l\eqno(3.36)$$
and then taking the limit $c\to \infty$ in (3.36).
For Cartan's 1--forms and the action the contraction procedure involves
rescaling the target superspace coordinates
$$\tilde\t^{i\a}=c\t^{i\a},\
\tilde x^{j\m} =c^2 x^{j\m}\eqno(3.36a)$$
and passing to the generators $P_\mu$, $Q_\alpha$.
One gets
$$\o^{j\m}_a={1\over c^2}{i\over 2}(\p_a\tilde x^{j\m}-
(-)^j i\p_a\tilde\t^j\G^\m\tilde\t^j)+0({1\over c^4})\ ,$$
$$\o^{j\a}_a ={i\over c}[\p_a\tilde\t^{j\a}-{i\over 4c^2}
(\p_a\tilde\t^{j\b}\tilde x^{j\m}-\p_a\tilde x^{j\m}
\tilde\t^{j\b}+$$
$$+(-)^j{i\over 3}(\p_a\tilde\t^j\Gamma^\m\tilde\t^j)
\tilde\t^{j\b})C^\a_{\m\b}]+0(1/c^5)\ . \eqno(3.37)$$
Substituting these expressions into (3.33) and keeping in the
 action the leading terms in ${1\over c}$ $(\sim {1\over c^4})$
 it is straightforward to
find $(\tilde x^\mu =\tilde x^{1\mu}-\tilde x^{2\mu})$
$$A={(-1)\over 4c^4} \int_{\p V}
d^2\xi \{-\ell_I \sqrt{-g} g^{ab}
(\p_a\tilde x^\m +i\sum^2_{j=1}
\p_a\tilde\t^j\G^\m\tilde\t^j)(\p_b
\tilde x_\m +i\sum^2_{j=1}
\p_b\tilde\t^j\Gamma_\m \tilde\t^j)-$$
$$-3\ell_{II}\v^{ab}
[i\p_a\tilde x^\m -\half \sum^2_{j=1}
\p_a\tilde\t^j\G^\m\tilde\t^j]
(\sum^2_{k,j=1}
\p_b\tilde\t^k\s^{kj}_3\G_\m\tilde\t^j)\}+0({1\over c^6})\eqno(3.38)$$
Comparing (3.38) with the GS superstring type action \cite{[10]} (2.18) one concludes that they
coincide (up to an inessential numerical coefficient $(-1)/4c^4$) iff
$$\ell_{II} =-{2\over 3}\ell_I\eqno(3.39)$$
(in case of  $\Gamma^\mu_{\alpha\beta}$ being
Dirac $\gamma$--matrices, one gets the genuine $N=2$
 GS covariant action (2.18)).

At this step it is appropriate to explain why for constructing
a nonabelian superstring action one has to start with the
self--dual supergroup $G = G_1\otimes G_2$ rather than
the product of two
isomorphic supergroups. The second option corresponds to the
 substitution
$\theta^{2\alpha} \to i\theta^{2\alpha}$ in (3.38), so in the
flat
superspace limit one would have, e.g.
$\partial_a\theta^1\Gamma^\mu\theta^1-\partial_a\theta^2\Gamma^\mu\theta^2$ instead of the correct GS expression
$\partial_a\theta^1\Gamma^\mu\theta^1+\partial_a\theta^2\Gamma^\mu\theta^2$.
Such a noncompact version of N=2 GS superstring (it is based on
$N=2$ supersymmetry with the $SO(1,1)$
automorphism group) would contain ghost fermionic degrees of freedom in the light--cone gauge.

The final expression for the invariant action of our $G_1\otimes G_2/G_+$ WZW sigma model is
as follows
$$A=\ell_I\{\int_{\p V} d^2\xi[-\sqrt{-g} g^{ab}
\o^{-\m}_a\o^-_{b\m}
+2\v^{ab}(\o^{1\m}_a
\o^2_{b\m})]-$$
$$-{2\over 3} \int_V d^3\xi\v^{ABC}[\p_A\o^{1\m}_B\o^1_{C\m}-\p_A\o^{2\m}_B\o^2_{C\m}+$$
$$+(\o^{1\a}_A\p_B\o^{1\b}_C+\o^{2\a}_A\p_B\o^{2\b}_C)X_{\ab}]\}\eqno(3.40)$$
Remarkably, when all the fermionic fields are put equal to zero,
(3.40) is reduced to the familiar conformally invariant WZW action for the even subgroup
$G$ (this is seen most directly when substituting the relation (3.39) into the action written in
the form (3.33)).
Thus, (3.40) can be regarded as a genuine $N=2$ superstring extension of the group manifold string action.

\subsection{The equations of motion.
Zero-curvature representation}

Being written through the covariant variations
$$\tilde\o^j=U^{-1}_j\d U_j =\tilde\o^{j\m}R^j_\m+\tilde\o^{j\a}S^j_\a\eqno(3.41)$$
$$\d\o^j_a-\p_a\tilde\o^j=[\o^j_a,\tilde\o^j]\eqno(3.42)$$
the full variation of the action (3.40) is given by
$$\d A =\ell_I\int_{\p V} d^2\xi[-\d(\sqrt{-g} g^{ab})
\o^{-\m}_a\o^-_{b\m}+2(\p_a(P^{ab}_-
\o^{1\m}_b-P^{ab}_+\o^{2\m}_b)-$$
$$-P^{ab}_+ \o^{1\l}_a\o^{2\r}_bt^\m_{\l\r})
\tilde\o^-_\m -2P^{ab}_+
\o^-_{b\m}
(\o^{1\a}_a\G^\m_{\ab}\tilde\o^{1\b})-$$
$$-2P^{ab}_-\o^-_{b\m}(\o^{2\a}_a\G^\mu_{\ab}
\tilde\o^{2\b})]\ ,\eqno(3.43)$$
where $\tilde\omega_\mu = \tilde\omega^1_\mu -\tilde\omega^2_\mu$ and
$P^{ab}_\pm =\sqrt{-g} g^{ab} \pm \varepsilon^{ab}$
\cite{[6]}.
In deducing (3.43), we have essentially used the equations (3.15), (3.42) and Jacobi
 identities for the structure constants.

The equations of motion following from (3.43) are written as
$$\p_a(P^{ab}_-\o^{1\mu}_b -P^{ab}_+
\o^{2\mu}_b) -t^\mu_{\l\r}P^{ab}_+\o^{1\l}_a\o^{2\r}_b =
0\eqno(3.44)$$
$$P^{ab}_+\o^{1\b}_a\G^\m_{\ab}\o^-_{b\mu}=0\ ,
P^{ab}_-\o^{2\b}_a\G^\mu_{\b\a}\o^-_{b\mu}=0\ ,
\eqno(3.45)$$
$$\o^{-\mu}_a\o^-_{b\mu}-\half g_{ab} g^{cd}\o^{-\m}_c
\o^-_{d\m}=0\eqno(3.46)$$
The last equation is obtained by varying the world--sheet
metric $g^{ab}$ and is nothing
else than the classical version of standard Virasoro constraints. Equation (3.44) is the equation
of motion for
$x^\mu(\xi)$  while Eqs. (3.45) are those for fermions
$\theta^{1\alpha}(\xi),\theta^{2\alpha}(\xi)$.
For completeness, one should also add to (3.44)--(3.45)
the MC equations (3.15).
Then this extended system gives a closed set of equations for the $G_1\otimes G_2/G_+$ Cartan 1--forms.
So, in our approach the latter can be treated as the primary independent objects
(cf. \cite{[14],[15]}, where analogous
equations entirely in terms of Cartan's forms have been obtained for ordinary bosonic sigma models,
including those with the WZ terms \cite{[15]}).

With exploiting the MC equations, the set (3.44), (3.45) can be given different equivalent representations.
One may, e.g. rewrite it as the two systems
$$(\nabla^1_a)^\mu_\lambda(P^{ab}_+\o^{-\l}_b)+\v^{ab}
\o^{1\a}_a\G^\m_{\ab}
\o^{1\b}_b=0\ ,$$
$$P^{ab}_+\o^-_{b\m}\G^\m_{\ab}\o^{1\b}_a=0\eqno(3.47a)$$
$$(\nabla^2_a)^\mu_\lambda(P^{ab}_-\o^{-\lambda}_b)-\v^{ab}\o^{2\a}_a\G^\m_{\ab}\o^{2\b}_b=0\ ,$$
$$P^{ab}_-\o^-_{\b\mu}\G^\m_{\ab} \o^{2\b}_a=0\ ,\eqno(3.47b)$$
where $(\nabla^j_a)^\mu_\lambda =\delta^\mu_\lambda \partial_a
+t^\mu_{\nu\lambda}\omega^{j\nu}_a$ are covariant derivatives with respect to gauge transformations (3.13).
Remarkably, like in the GS superstring case (Eqs.(2.23)) one may cast (3.47) into a simple form as the conservation
laws for the appropriate linearly transforming currents
(see Eq.(3.22)).
$$\p_a(P^{ab}_+J^{L-}_b +\v^{ab}J^{L1}_b)=0\ ,\eqno(3.47a')$$
$$\p_a(P^{ab}_-J^{R-}_b-i\v^{ab}J^{R2}_b)=0\eqno(3.47b')$$
Just these combinations of currents generate global $G_1$ and $G_2$ symmetries of the action (3.40).

The property that the equations of motion are divided into the two sets, respectively for the ``left''
and ``right'' variables
$\{P^{ab}_+\omega^{-\lambda}_b,\omega^{1\alpha}_a\}$,
$\{ P^{ab}_-\o^{-\lambda}_b,\o^{2\a}_a\}$
seems to reflect the product structure of the underlying
supergroup $G=G_1\otimes G_2$.
So we expect $G_1$ to be eventually realized on the
left variables while $G_2$  on the right ones.
Of course, this statement can be given a precise meaning
only on the solutions of these equations, e.g. after passing to a light--cone type gauge, and upon performing
quantization.

A remarkable property of the obtained equations is their
complete integrability.

A zero curvature representation for Eqs.(3.47) is written as
$$[L^+_a,L^+_b]=[L^-_a,L^-_b]=0\eqno(3.48)$$
where
$$L^+_a=\p_a-\half(1-\l^2)\v_{ab}P^{bc}_+J^{L-}_c-(\l+1)J^{L1}_a$$
$$L^-_a=\p_a-\half(1-\l^2)\v_{ab}P^{bc}_-J^{R-}_c+i(\l+1) J^{R2}_a\eqno(3.49)$$
and the currents
$J^{L-}_c,\ J^{R-}_c,\ J^{L1}_a,\
J^{R2}_a$ have been defined in Eqs.(3.22).
Here, the generators $R_\mu, S_\alpha$ satisfy the superalgebra (3.1);
$\lambda$ is a spectral parameter.

The vanishing of the first and second commutators in (3.48) yields, respectively,
Eqs.(3.47a) and (3.47b)
(to be more precise, their current form (3.47a$'$) and (3.47b$'$)).
The MC  equations (3.15) are also encoded in the
integrability conditions (3.48).
Note that $L^+_a$ is in fact defined on the super--algebra (3.1) while $L^-_a$ on (3.3) (one could define $L^+_a, L^-_a$ on
the two sets of
mutually commuting abstract generators $R^1_\mu$,
$S^1_\alpha$ and $R^2_\mu,\ S^2_\alpha$ obeying (3.1) and (3.3);
in (3.49) for simplicity we have related $L^+_a, L^-_a$ to the same set of generators).

The possibility to represent the equations of motion in the form (3.48),
(3.49) leads us to the conclusion that we have constructed a new
completely integrable 2D system.
Thus, the integrability of the abelian $N=2$ GS superstring
type models (including genuine $N=2$ GS superstring)
mentioned in Sec.1 turns out a particular  case of a more general phenomenon
inherent in the nonabelian models. As usual, the integrability suggests the existence
of infinitely many conserved currents which can be evaluated using
an auxiliary spectral problem associated with the operators
$L^\pm_1$ (see e.g. \cite{[16]}).
Also, this is an indication that the set of Eqs. (3.47) can be linearized in proper variables. Let
us emphasize a crucial role of the WZ term for attaining the integrability. We have checked that
without this term the equations of motion can never be brought into the form (3.48). Moreover, the
integrability comes out only if the free parameters $\ell_{I}$ and $\ell_{II}$ are adjusted so as
to give the action (3.40). This is in accordance with the general statement of \cite{[17]} about
nonintegrability of ordinary (i.e. having no WZ terms) sigma models defined on nonsymmetric coset
spaces. Our consideration demonstrates that in a number of cases the integrability can be achieved
by adding proper WZ terms to the sigma model action. It seems that this phenomenon is not specific
for supergroups only. It would be interesting to study its implications in purely bosonic 2D sigma
models.

For completeness, we present the  equations of motion
(3.44), (3.45) and the MC equations (3.15) in the conformal gauge
$$\sqrt{-g} g^{ab} =\left({1\atop 0}\ {0\atop -1}\right),\
\partial_\pm =\half(\partial_0\pm\partial_1),\
\o_\pm=\half(\o_0\pm\o_1),$$
$$\p_+\o^{1\m}_--\p_-\o^{2\m}_+ -t^\m_{\l\r}
\o^{1\l}_-\o^{2\r}_+=0\ ,$$
$$\G^\m_{\ab}\o^-_{+\m}\o^{1\a}_-=0\ ,$$
$$\G^\m_{\ab}\o^-_{-\m}\o^{2\a}_+ = 0 \eqno(3.50)$$
$$\p_+\o^{1\m}_--\p_-\o^{1\m}_+ +t^\m_{\n\l}
\o^{1\n}_+\o^{1\lambda}_-+\G^\m_{\ab}\o^{1\a}_+\o^{1\b}_-=0\ ,$$
$$\p_+\o^{2\m}_--\p_-\o^{2\m}_++t^\m_{\n\l}\o^{2\n}_+\o^{2\l}_-
-\G^\m_{\ab}\o^{2\a}_+\o^{2\b}_-=0\ ,$$
$$\p_+\o^{j\a}_--\p_-\o^{j\a}_++C^\a_{\m\b}
(\o^{j\m}_+\o^{j\b}_--\o^{j\b}_+\o^{j\m}_-)=0\eqno(3.51)$$

\subsection{$\kappa$--supersymmetry}

Our last topic will be exploring the conditions under which the action (3.40) possesses a local fermionic
$\kappa$--symmetry \cite{[6],[18]}.

The only additional requirement needed for ensuring such a symmetry is the existence of matrices $(\tilde\Gamma^\mu)^{\beta\gamma}$
satisfying the relations
$$\G^\m_{\ab}(\tilde\G^\nu)^{\b\g}+\G^\n_{\ab}(\tilde\G^\m)^{\b\g}
=2\eta^{\m\n}\rho_\a^\g\ ,\eqno(3.52)$$
where $\rho^\gamma_\alpha$ is an arbitrary matrix which can be degenerate.

Using the general expression for $\delta A$ (3.43) it is easy to show that $\delta A$ vanishes under the following
variations
$$\tilde\o^{-\m}=0$$
$$\tilde\o^{1\a}=P^{ab}_+\o^-_{b\mu}(\tilde\G^\m)^{\ab}
\kappa^1_{a\b}(\xi)\ ,$$
$$\tilde\o^{2\a}=P^{ab}_-\o^-_{b\mu}(\tilde\G^\m)^{\ab}
{\kappa}^2_{a\b}(\xi)$$
$$\delta(\sqrt{-g} g^{ab})=-2(P^{da}_+
P^{cd}_+(\o^{1\b}_d\r^\a_\b{\kappa}^1_{c\a})
-P^{da}_-P^{cb}_-(\o^{2\b}_d\rho^\a_\b{\kappa}^2_{c\a}))
\eqno(3.53)$$
Here ${\kappa}^{1,2}_{a\beta}(\xi)$ are the odd transformation
 parameters. These
transformations go over to the standard
${\kappa}$--symmetry transforamtions of the ordinary
$N=2$ GS superstring in the flat superspace limit
(when $\Gamma^\mu_{\alpha\beta}C^{\beta\gamma}$ coincide with Dirac $\gamma$ matrices)
and thus generalize this symmetry to nonabelian case. Like in the case of the GS superstring model,
due to the presence of local ${\kappa}$--sypersymmetry, the theory
 in question is expected to have a representation via free fields
(in a suitable gauge). Thus, if Eq.(3.52) has a solution, the nonabelian superstring model seems to be reducible to a 2D conformal field theory.
Otherwise, local ${\kappa}$--supersymmetry is lacking and an equivalence of the full action (3.40) to a conformally--invariant one becomes questionable
(though conformal invariance still persists for the bosonic part of (3.40)).
Fortunately, even in this case our model is expected to admit a kind of linearization because of its complete
integrability. Whether this possibility is still related to some
infinite--dimensional (local) supersymmetry hidden in the zero curvature representation (3.48)
is an interesting open question.

Finally, let us emphasize that the condition (3.52) places severe limitations on the
appropriate class of superalgebras (3.1).
It is easy to see that, in the case of a nondegenerate matrix
$\rho$, Eq.(3.52) is the defining relation of some Clifford algebra.
Indeed, (3.52) implies that $\tilde\Gamma^\nu \rho^{-1}$ like $\Gamma^\nu$ is a symmetric
matrix. So, $\Gamma^\nu$ and $\tilde\Gamma^\nu$ can be combined into a single matrix
\footnote{We thank P.S. Howe for suggesting this to us.}
$$\g^\m =\left({0\atop\tilde\G^\m\rho^{-1}}\
{\Gamma^\mu\atop 0}\right)\eqno(3.54)$$
which satisfies, as a consequence of (3.52), the standard Clifford algebra relation
$$\g^\m\g^\n+\g^\n\g^\m = 2\eta^{\m\n}\eqno(3.55)$$
$(\Gamma^\mu$ and $\tilde\Gamma^\mu\rho^{-1}$ are a kind of Weyl projections of $\gamma^\mu$).
If $\Gamma^\m_{\a\b}$ provides an irreducible
representation of (3.55), we have severe
restrictions on the dimensions of generators entering into
the superalgebra (3.1).
Otherwise, $\Gamma^\mu_{\alpha\beta}$ have a block structure,
each block
corresponding to an irredicible representation of Clifford algebra
(3.55).
In view of the general cyclic identity (3.2), for each block one also has the well--known
restrictive relations between the dimensions of vector and spinor indices. In the case of a
degenerate matrix $\rho$ we are led to reduce the space on which this matrix acts, after which we
are again left with the nondegenerate situation discussed above. An example of the superaglebra
with a nondegenerate metric for which one may find $\tilde\Gamma^\nu$ (3.52) corresponding to a
degenerate matrix $\rho$ is given by the Green superalgebra \cite{[12]}. It can be cast into the
form (3.1) by joining its supertranslation generator $Q_\alpha$ and the extra fermionic central
charge generator $K^\alpha$ into a single generator
$S_{\hat\alpha}$.
For this case we have
$$\Gamma^\mu = \tilde\gamma^\mu \sim \left({\Gamma^\mu_1
\atop 0}\quad {0\atop 0}\right)$$
and $\Gamma^\mu_1$ are matrices which are related to
 Dirac $\gamma$ matrices
and satisfy the identity (3.2).
It would be interesting to inquire whether there exist nontrivial
superaglebras (3.1) of that sort having a nonabelian even part.

\setcounter{equation}{0}
\setcounter{section}{4}

\section*{4 $\;\;\;$ EXAMPLES \footnote{This Section (together with refs. [21] - [23]) is absent in the original ICTP preprint. It
has been prepared for a revised version of the paper submitted
to a journal. We expose it here also without updating.}}

Here we illustrate the general consideration of previous Sections
by simple examples of nonabelian $N=2$ superstrings. All these
examples reveal local fermionic $\kappa$-invariance, thus
demonstrating that the variety of the $\kappa$-invariant
nonabelian $N=2$ superstring models is not empty.

Actually, listing all possible models of that sort amounts to
classifying all possible superalgebras (3.1) which have a
non-degenerate Killing supermetric and whose structure constants
$\Gamma^{\mu}_{\alpha\beta}$ obey the Clifford algebra condition
(3.52). In other words, one needs to classify the superalgebras
with the structure constants $\Gamma^{\mu}_{\alpha\beta}$ forming
a representation (irreducible or reducible) of Clifford algebra.
We have no a general solution to this interesting algebraic
problem as yet. However, now we are aware of a number of
particular examples of such superalgebras which can be used to
construct new non-trivial superstring type models.

The list of $\kappa$-invariant models known to us involves those associated with the superalgebras
$osp(2|1)$, $su(2|1)$, as well as with various direct sums of these superalgebras which may in
addition include the Green superalgebra \cite{[12]}. As was already mentioned, the latter also
belongs to the type we are interested in.

In order to be able to relate the considered models to special solutions of a curved background, it
is advantageous to rewrite the general action (3.40) as the action for the superstring moving in a
$N=2$ supergravity background \cite{[19]}
\begin{eqnarray}
A &=& l_{I} \left\{ \int_{\partial V} d^{2}\xi
\left[ -\frac{1}{2}\Phi \left( z(\xi) \right)
\eta_{\mu\nu}\sqrt{-g} g^{ab}\left( \partial_{a} z^{M} E_{M}^{\mu} \right)
\left( \partial_{b} z^{N}E_{N}^{\nu} \right) \right] + \right. \nonumber \\
&+& \left. \frac{1}{2} \int_{V} d^{3}\xi
\varepsilon^{abc}\partial_{a} z^{M} \partial_{b} z^{N}
\partial_{c} z^{K} H_{MNK} \right\}.
\end{eqnarray}
Here $\Phi(z)$ is the dilaton superfield which appears in the
superfield formulations of the type II supergravities (one can get
rid of this superfield by rescaling $E_{M}^{\mu} \rightarrow
\frac{1}{\sqrt{\Phi}}E_{M}^{\mu}$), $z^{M}=\{ x^{\mu},
\theta^{1\alpha}, \theta^{2\alpha} \}$ are coordinates of the
coset space $G_{1} \times G_{2} /G_{+}$ (we fix the right gauge
freedom by the condition $x^{1\mu} + x^{2\mu} = 0$ and thus put
$x^{\mu} = \frac{1}{2}\left( x^{1\mu} - x^{2\mu} \right) =
x^{1\mu} = -x^{2\mu}$) and $E_{M}^{\mu} =
\frac{1}{\sqrt{\Phi}}\left( E_{M}^{1\mu} - E_{M}^{2\mu} \right)$
where $E_{M}^{j\mu}$ are related to the previously used quantities
as
$$\omega_{a}^{j\mu} = \partial_{a} z^{M}E_{M}^{j\mu}.$$
Finally, the three-index field strength $H_{MNK}$ is related to
the three-form (see eq.(3.33))
\begin{eqnarray}
H &=& dz^{M}dz^{N}dz^{K} H_{MNK} \;=\; \nonumber \\
  &=& \frac{1}{3}\Phi^{3/2}t_{\mu\nu\lambda}E^{\mu}E^{\nu}E^{\lambda} + \Phi^{1/2}\Gamma_{\mu,\alpha\beta}\left( E^{\mu}E^{1\alpha}E^{1\beta} - E^{\mu}E^{2\alpha}E^{2\beta} \right),
\end{eqnarray}
where
\begin{eqnarray}
t_{\mu\nu\lambda} &=& -\eta_{\mu\rho}t_{\nu\lambda}^{\rho},\;
\Gamma_{\mu,\alpha\beta} \;=\;
\eta_{\mu\nu}\Gamma_{\alpha\beta}^{\nu}, \; E^{A}\;=\;\left(
\Phi^{-1/2}\left( E^{1\mu} - E^{2\mu} \right), \;
E^{1\alpha}, \;E^{2\alpha} \right) \nonumber \\
E^{j\mu} &=& dz^{M}E_{M}^{j\mu},\; \;
E^{j\alpha}\;=\;dz^{M}E_{M}^{j\alpha} \equiv
d\xi^{a}\omega_{a}^{j\alpha}.
\end{eqnarray}

Let us now turn to a brief description of our examples.
\subsection{The $OSp(2|1, R)$ model}
The superalgebra $osp(2|1,R)$ includes three even and two odd
generators which obey the (anti)commutation relations of the form
(3.1) with
$$C_{\mu}\;=\;\left\{ -\frac{i}{2}\sigma_{2},\;\frac{1}{2}\sigma_{3},\;\frac{1}{2}\sigma_{1} \right\}\;=\;\Gamma_{\mu}X \eqno{(4.4a)}$$
$$\Gamma^{\mu}\;=\;\left\{ \frac{i}{2}
\left(
\begin{array}{cc}
-1  &  0 \\
 0  & -1
\end{array}
\right), \frac{i}{2} \left(
\begin{array}{cc}
0 & 1 \\
1 & 0
\end{array}
\right), \frac{i}{2} \left(
\begin{array}{cc}
-1 & 0 \\
 0 & 1
\end{array}
\right) \right\},\;
X^{\alpha\beta}\;=\;i\varepsilon^{\alpha\beta},\;\varepsilon^{12}\;=\;1
\eqno{(4.4b)}$$
$$\eta_{\mu\nu}\;=\;-Str\left( R_{\mu}R_{\nu} \right) = diag\left( 1,-1,-1 \right),\;\; t_{\mu\nu\rho}\;=\;-\eta_{\mu\lambda}t_{\nu\rho}^{\lambda}\;=\;\varepsilon_{\mu\nu\rho}. \eqno{(4.4c)}$$
\setcounter{equation}{4} Here $\sigma_{\mu}$ are Pauli matrices
and $\varepsilon_{123}=1$. Note that the even subalgebra of
$osp(2|1,R)$ is $sl(2,R)\sim so(1,2)$.

Due to the property that $\Gamma^{\mu}$ are related to Pauli
matrices, it is easy to construct the matrices
$\tilde{\Gamma}^{\mu}$ satisfying eq.(3.52) with $\rho = 1$
\begin{equation}
\tilde{\Gamma}^{\mu}\;=\;\left\{ 2i \left(
\begin{array}{cc}
1  &  0 \\
 0  & 1
\end{array}
\right), 2i \left(
\begin{array}{cc}
 0 & 1 \\
1 &  0
\end{array}
\right), 2i \left(
\begin{array}{cc}
 -1  &  0 \\
  0  &  1
\end{array}
\right) \right\}.
\end{equation}
Thus, the nonabelian sigma model (4.1) with the target superspace
$$OSp(2|1)\times OSp(2|1)^{*}/SL(2,R)_{diag}$$
possesses fermionic $\kappa$-invariance and hence can be
interpreted as the model of $N=2$ superstring moving in a special
$D=3,\;N=2$ supergravity background. Comparing the particular
value of the field strength $H_{MNK}$ in (4.2) for the choice
(4.4c) with the general expression for $H_{MNK}$ in $D=3,\;N=1$
Poincare supergravity \cite{[20a]}, we find that the nonvanishing tangent space coefficients of
this strength are the same in both cases. This coincidence leads us to the conclusion that the
above background supplies an $N=2$ extension of some particular solution to $D=3,\;N=1$ Poincare
supergravity \cite{[21a]}, with $OSp(2|1)$ as the isometry supergroup.

Being $\kappa$-invariant, the model under consideration is
expected to be conformally-invariant and to possess world-sheet
supersymmetry. The latter can be argued by counting essential
bosonic and fermionic degrees of freedom. After imposing the
light-cone gauge there remain $2(D-2)=2$ bosonic degrees, one left
and one right movers. Accordingly, two of four originally present
fermionic degrees of freedom are gauged away by $\kappa$-symmetry,
so that the final number of fermionic coordinates equals 2. Thus,
the number of unremovable bosonic variables matches with that of
fermionic ones, which is a strong indication of hidden 2D
world-sheet supersymmetry. In turn this suggests that the
superstring model we deal with is equivalent to some fermionic
string model.

\subsection{The $SU(2|1)$ model}
The superalgebra $su(2|1)$ includes four even generators
($R_{\mu}, \; \mu\;=\;0,1,2,3$) which form the algebra $u(2) =
u(1) + su(2)$ and four odd generators (a complex $SU(2)$ doublet
$\Gamma_{\hat{\alpha}},\;\;\;
\overline{\Gamma}^{\hat{\beta}}\;\equiv \;\left(
\Gamma_{\hat{\beta}} \right)^{\dagger}$;
$\hat{\alpha},\;\hat{\beta},...=1,2$).

The (anti)commutation relations of this superalgebra can be found in \cite{[22a]}. We will be
interested here mainly in the anticommutation relations which can be written as
\begin{equation}
\{ \Gamma_{\hat{\alpha}}, \overline{\Gamma}^{\hat{\beta}}
\}\;=\;\left( \tau^{\mu} \right)_{\hat{\alpha}}^{\hat{\beta}}
R_{\mu},\;\; \{ \Gamma_{\hat{\alpha}}, \Gamma_{\hat{\beta}}
\}\;=\;\{ \overline{\Gamma}^{\hat{\alpha}},
\overline{\Gamma}^{\hat{\beta}} \} \;=\;0,
\end{equation}
where
\begin{equation}
\left( \tau^{\mu} \right)_{\hat{\alpha}}^{\hat{\beta}}\;=\;\left\{
\delta_{\hat{\alpha}}^{\hat{\beta}}, \; \left( \sigma_{1}
\right)_{\hat{\alpha}}^{\hat{\beta}},\; \left( \sigma_{2}
\right)_{\hat{\alpha}}^{\hat{\beta}},\;\left( \sigma_{3}
\right)_{\hat{\alpha}}^{\hat{\beta}} \right\}. \nonumber
\end{equation}
Combining $\Gamma$ and $\overline{\Gamma}$ into a single
self-conjugated spinor $S_{\alpha}\;=\; (
\Gamma_{\hat{\alpha}},\;\; \overline{\Gamma}^{\hat{\beta}} )$
($\alpha,\beta,...=1,2,3,4$) we represent eqs.(4.6) in the form
\begin{equation}
\{ S_{\alpha}, S_{\beta} \}\;=\;\Gamma_{\alpha\beta}^{\mu}R_{\mu},
\end{equation}
where (the superscript $T$ means matrix transposition)
\begin{equation}
\Gamma^{\mu} \;=\; \left\{ \left(
\begin{array}{cc}
 0  &  I    \\
 I  &  0
\end{array}
\right), \; \left(
\begin{array}{cc}
            0   & \sigma_{1}  \\
\sigma_{1}^{T}  &  0
\end{array}
\right), \; \left(
\begin{array}{cc}
             0  & \sigma_{2} \\
 \sigma_{2}^{T} &  0
\end{array}
\right),\; \left(
\begin{array}{cc}
             0  & \sigma_{3} \\
\sigma_{3}^{T}  &  0
\end{array}
\right) \; \right\}.
\end{equation}
Now the whole set of the (anti)commutation relations of $su(2|1)$
can be given the generic form (3.1) with
$$
t_{ik}^{l} = \varepsilon_{ikl},\;\;t_{0\mu}^{\lambda} =
t_{\mu\nu}^{0} = 0,\;\; X = \left(
\begin{array}{cc}
 0 & I \\
-I & 0
\end{array}
\right).$$

It is straightforward to find the matrices $\tilde{\Gamma}^{\mu}$
which satisfy eq.(3.52) with $\eta_{\mu\nu}\;=\;-Str\left(
R_{\mu}^{ad}R_{\nu}^{ad} \right)\;=\;diag(1,-1,-1,-1)$
\begin{equation}
\tilde{\Gamma}^{\mu}\;=\; \left\{ \left(
\begin{array}{cc}
 0  &  I    \\
 I  &  0
\end{array}
\right), \; \left(
\begin{array}{cc}
           0  & -\sigma_{1}^{T}  \\
 -\sigma_{1}  &  0
\end{array}
\right), \; \left(
\begin{array}{cc}
           0 & -\sigma_{2}^{T} \\
 -\sigma_{2} &  0
\end{array}
\right),\; \left(
\begin{array}{cc}
          0  & -\sigma_{3}^{T} \\
-\sigma_{3}  &  0
\end{array}
\right) \; \right\}.
\end{equation}

Let us remark that the even $su(2|1)$ generators in adjoint
representation $R_{\mu}^{ad}$ are given by the following
$6\times6$ supermatrices
$$
R_{0}^{ad} = \frac{1}{2} \left(
\begin{array}{c|c}
0 & 0 \\
\hline 0 & \begin{array}{cc}
-I & 0 \\
 0 & I
\end{array} \end{array}
\right), \; R_{j}^{ad} = \frac{1}{2} \left(
\begin{array}{c|c|c}
0 & 0 & 0 \\
\hline
0 & -2i\varepsilon_{jkl} & 0 \\
\hline 0 & 0 & \begin{array}{cc}
\sigma_{j} & 0 \\
         0 & -\sigma_{j}
\end{array} \end{array}
\right), j,k,l = 1,2,3
$$
and we meet here an interesting situation when the even subgroup
of $SU(2|1)$ is compact, $SU(2) \times U(1)$, while the bosonic
part of Killing supermetric is pseudo-Euclidean,
$\eta_{\mu\nu}\;=\;-t^{\lambda}_{\mu\rho} t^{\rho}{\nu\lambda} +
C^{\beta}_{\mu\alpha} C^{\alpha}_{\nu\beta}\;=\;-Str\left(
R_{\mu}^{ad}R_{\nu}^{ad} \right)\;=\;diag\left( 1,-1-1-1 \right)$.
Of course, the latter by no means contradicts the compactness of
$SU(2) \times U(1)$, since this symmetry, in its own right, does
not fix the relative sign of $\eta_{00}$ and $\eta_{ij}$. It is
remarkable that the same indefinite signature of $\eta_{\mu\nu}$
is required for the existence of the matrix $\tilde{\Gamma}^{\mu}$
and hence for $\kappa$-invariance (no appropriate matrices
$\tilde{\Gamma}^{\mu}$ exist with the choice
$\eta_{\mu\nu}\;=\;\pm diag\left( 1,1,1,1 \right)$).

Thus, the superstring sigma model associated with the coset
$$SU(2|1) \times SU(2|1)^{*}/U(2)_{diag}$$
also respects $\kappa$-supersymmetry. This model can probably be
interpreted as describing a superstring moving in a special $D=4,
N=2$ supergravity background with the isometry supergroup $SU(2|1)
\times SU(2|1)^{*}$ and $S^{1} \times S^{3}$ (or $R^{1} \times
S^{3}$) as the bosonic submanifold\footnote{The $U(1)$ generator
included in the coset space is realized as shifts of the
coordinate $x^{0}(\xi)$, so we may treat the abelian factor in the
bosonic submanifold to be compact $S^{1}$ or noncompact $R^{1}$
depending on the boundary conditions imposed on $x^{0}$.}. Once
again, counting physical bosonic and fermionic degrees of freedom
indicates the potential presence of a world-sheet supersymmetry.
Indeed, the number of physical bosonic coordinates equals
$2(D-2)=4$ that coincides with the net number of fermionic
coordinates which remain after utilizing $\kappa$-invariance
($\frac{1}{2}(4 + 4) = 4$).

\subsection{Models based on direct sums of superalgebras}
For any sum of the superalgebras considered above the identity
(3.52) is obviously satisfied and the relevant $\Gamma^{\mu}$ and
$\tilde{\Gamma}^{\mu}$ form reducible representations of Clifford
algebra. As a result, the corresponding superstring sigma models
are guaranteed to be $\kappa$-invariant. Here we describe a class
of such models, with the bosonic manifolds being products of a
flat Minkowski space and some nonabelian group manifolds. Many of
these models reveal equal numbers of physical bosonic and
fermionic coordinates and so have a chance to be world-sheet
supersymmetric (and hence equivalent to some fermionic string
models).

This particular class of the $G \times G^{*}/G_{+}$ sigma models
is associated with the non-semisimple supergroups $G$ generated by
the following direct sums of superalgebras
$${\cal G}\;=\;{\cal G}_{G} + \sum_{1}^{L} su(2|1) + \sum_{1}^{K} osp(2|1), \eqno{(4.11)}$$
\setcounter{equation}{11}
where ${\cal G}_{G}$ is the Green algebra \cite{[12]}. The flat Minkowski space part of the
corresponding bosonic coordinates is related to the translation generators in
${\cal G}_{G}$ \footnote{To avoid a misunderstanding, we point out
that the Minkowski space Lorentz group is assumed to be placed in
the stability subgroup.} while the rest of these coordinates is
valued in products of the $U(2)$ and $SL(2,R)$ group manifolds.
The bosonic tangent space metric in the present case can be chosen
in the form $\eta_{\mu\nu}\;=\;diag\left(1,\;-1,\;-1,\ldots
1,\;-1,\;-1,\ldots 1,\;-1,\;-1,\ldots -1,\;-1 \right)$, first $+1$
coming from the flat Minkowski metric and the remaining $K + L$
ones from the $SL(2,R)$ and $U(1)$ factors.

Let us deduce the conditions under which these models possess
equal numbers of physical bosonic and fermionic coordinates.
Denoting by $D$ the full dimension of the bosonic manifold, we
have
\begin{equation}
D\;=\;d+3K+4L,
\end{equation}
$d$ being the dimension of Minkowski space. Thus, in the
light-cone gauge one is left with
\begin{equation}
N_{b}\;=\;2(D-2)\;=\;2(d+3K+4L-2)
\end{equation}
essential bosonic degrees of freedom ($D - 2$ left and $D - 2$
right movers). Now recall that Green's superalgebras exist only in
dimensions $d\;=\;3,4,6,10$ and have, respectively, $4,8,16$ and
$32$ real fermionic generators \cite{[12]} (two Majorana spinors for
$d\;=\;3,4$, two complex Weyl spinors for $d\;=\;6$ and two
Majorana-Weyl spinors for $d\;=\;10$). Further, as was mentioned
in Sec.3.6, local $\kappa$-supersymmetry is capable to gauge away
half of the fermionic coordinates associated with one of two
spinor generators of the Green superalgebra, leaving intact the
coordinates related to the second generator. Then it is easy to
obtain that the Green supergroup sector of our $G \times
G^{*}/G_{+}$ sigma model contributes
\begin{equation}
N_{G}\;=\;2(3d - 6)
\end{equation}
of physical fermionic degrees of freedom. The total number of such
degrees equals
$$N_{f}\;=\;2(3d - 6 + K +2L). \eqno{(4.14)}$$
Finally, the condition
$$N_{b}\;=\;N_{f}$$
yields the simple necessary criterion for the existence of
world-sheet supersymmetry in the considered class of models
$$d - K - L\;=\;2. \eqno{(4.15)}$$
Substituting the admissible values $3,4,6,$ and $10$ of $d$ into
(4.15), we may figure out those values of $K,L$ and $D$
(eq.(4.12))\footnote{Note that the whole dimension of the bosonic
subspace, as it follows from (4.12), is summed up from the
``magic'' dimensions $3,4,6,10$ needed for the existence of the
closed Wess-Zumino three-forms having non-vanishing flat
superspace limit and for $\kappa$-supersymmetry.} which correspond
to the potentially world-sheet supersymmetric superstring models

The number of possible solutions rapidly increases with growing of
$d$, so we limit ourselves to presenting several noticeable
examples.

First, we quote the solutions corresponding to four-dimensional Minkowski space \\
\vskip6pt \underline{$d\;=\;4$}
$$1)\;K\;=\;0,\;\;L\;=\;2\;\;(D\;=\;12);\;\;\;2)\;K\;=\;1,\;\;L\;=\;1\;\;(D\;=\;11); \eqno{(4.16)}$$
$$3)\;K\;=\;2,\;\;L\;=\;0\;\;(D\;=\;10);$$
\setcounter{equation}{16} Third solution in (4.16) seems to be
most interesting because the related sigma model is expected to
represent the superstring moving in a special $D=10,\;N=2$
supergravity background which is reduced in the bosonic sector to
the product of four-dimensional Minkowski space and
six-dimensional non-compact group manifold $\sim SO(1,2) \times
SO(1,2)$.

Other interesting models arise at the choices
\begin{equation}
d\;=\;3,\;\;K\;=\;1,\;\;L\;=\;0\;\;(D\;=\;6)
\end{equation}
and
\begin{equation}
d\;=\;3,\;\;K\;=\;0,\;\;L\;=\;1\;\;(D\;=\;7)
\end{equation}
The model (4.17) can be interpreted as the superstring in a
$D=6,\;N=2$ supergravity background the bosonic part of which is
the product of three-dimensional Minkowski space and
three-dimensional non-compact group manifold $\sim SO(1,2)$. The
model characterized by the parameters (4.18) represents a
$D=7,\;N=2$ superstring moving in the background which is reduced
in the bosonic sector to the product of three-dimensional
Minkowski space and four-dimensional compact manifold $S^{1}
\times S^{3}$ related to the group $U(1) \times SU(2)$ (the circle
$S^{1}$ can be replaced by a real line $R^{1}$, see last
footnote).

We hope to return to studying these models elsewhere.

\section{CONCLUSION}

In this paper we have constructed the new WZW models on
supergroups which can be interpreted as models of $N=2$ GS
superstrings moving in supergroup spaces. In doing so, we
exploited the basic concepts of our reformulation of ordinary
$N=2$ GS superstring as a WZW sigma model on the product of two
N=1 supertranslation groups \cite{[9],[10]}. It was not clear before how to consistently construct
nonabelian versions of the GS $N=2$ superstring proceeding from its conventional sigma model
interpretation \cite{[7],[8]} or by choosing an appropriate supergravity background in the model of
a superstring moving in the general curved N=2 superspace \cite{[19]}. So, the main merit of our
formulation of ordinary $N=2$ GS superstring has to be seen in the possibility to straightforwardly
extend it to the nonabelian case.

The major idea of our approach is to consider a WZW sigma model
with the target coset space $G_1\otimes G_2/G_+$ where $G_1\otimes
G_2$ is a direct product of two supergroups dual to each other in
Cartan's sense and $G_+$ is the maximal even diagonal subgroup of
$G_1\otimes G_2$. We have shown that if $G_1$ and $G_2$ have a
flat limit coinciding with two N=1 supertranslation groups, then
our model in this limit coincides with the N=2 GS superstring
theory. We have deduced the equations of motion and proved their
 classical integrability (for arbitrary gauge fixing).
This is important for setting up a self--consistent quantum
version of our models which in general possess no
${\kappa}$--supersymmetry and for this reason may be not
conformally invariant.

We think that our construction can be extended to the superstring theories having nonequivalent
left and right sectors in the space of string variables (similarly to the heterotic string).
Besides, we expect that it can be directly applied to the case when $G_1$ and $G_2$ are arbitrary
$Z_2$ graded groups (not necessarily supergroups). Proceeding in this way, it seems possible to
construct new integrable sigma models and to formulate new string theories, perhaps, with the
critical dimensions close to the physical value of four. One more possible application of our
models is an unconventional way of introducing gauge degrees of freedom into string theories
without using Chan--Paton factors or additional fermions \cite{[20]}.

The important problems concerning the Hamiltonian formulation and
quantization of the present models as well as their connection
with $N=2$ superstring moving in an arbitrary supergravity
background will be addressed in our forthcoming paper.

\vskip2truecm \centerline{\bf Acknowledgements}

One of the authors (E.I.) would like to thank Professor Abdus
Salam, the International Atomic Energy Agency, and UNESCO for
hospitality at the International Centre for Theoretical Physics,
Trieste where this work was completed. He would also like to thank
Professor E. Sezgin for useful discussions.
\vfill\eject



\end{document}